\documentclass[aps,noshowpacs,english,superscriptaddress,twocolumn]{revtex4}
\usepackage{amsmath,amssymb,amsfonts,mathrsfs}
\usepackage{pgfplots,pgfplotstable,physics,epstopdf}
\usepackage{amsthm}
\usepackage{CJK}
\usepackage{bm}
\usepackage{babel}
\usepackage{graphicx}
\usetikzlibrary{shadings}
\usepackage[colorlinks=true,linkcolor=blue,anchorcolor=red,citecolor=blue,urlcolor=blue]{hyperref}
\usepackage{url,natbib,doi}
\usepackage{appendix}
\usepackage{tikz,pgf}
\usepackage{pdfpages}

\makeatletter
\AtBeginDocument{\let\LS@rot\@undefined}
\makeatother

\def \K {\hat{\mathcal{K}}}
\def \Z {\mathbb{Z}}
\def \H {\mathcal{H}}
\def \k {\bm{k}}

\def \TT {\hat{\mathcal{T}}}

\def \PP {\hat{\mathcal{P}}}

\def\T{\mathcal{T}}
\def\C{\mathcal{C}}
\def\P{\mathcal{P}}
\def\S{\mathcal{S}}

\def \Q {\mathcal{Q}}
\def \UU {\mathcal{U}}

\def\W{\mathcal{W}}

\def \I {\hat{I}}

\def\a{\mathbf{a}}
\begin{document}
	\renewcommand{\figurename}{FIG.}

\title{Takagi Topological Insulator on the Honeycomb Lattice}

\author{Qing Liu}
\email[These authors contributed  equally  to this work.]{}
\affiliation{National Laboratory of Solid State Microstructures and Department of Physics, Nanjing University, Nanjing 210093, China}

\author{Kai Wang}
\email[These authors contributed  equally  to this work.]{}
\affiliation{National Laboratory of Solid State Microstructures and Department of Physics, Nanjing University, Nanjing 210093, China}

\author{Jia-Xiao Dai}
\email[These authors contributed  equally  to this work.]{}
\affiliation{National Laboratory of Solid State Microstructures and Department of Physics, Nanjing University, Nanjing 210093, China}


\author{Y. X. Zhao}
\email[]{zhaoyx@nju.edu.cn}
\affiliation{National Laboratory of Solid State Microstructures and Department of Physics, Nanjing University, Nanjing 210093, China}
\affiliation{Collaborative Innovation Center of Advanced Microstructures, Nanjing University, Nanjing 210093, China}

\begin{abstract}
Recently, real topological phases protected by $PT$ symmetry have been actively investigated. In two dimensions, the corresponding topological invariant is the Stiefel-Whitney number. A recent theoretical advance is that in the presence of the sublattice symmetry, the Stiefel-Whitney number can be equivalently formulated in terms of Takagi's factorization. The topological invariant gives rise to a novel second-order topological insulator with odd $PT$-related pairs of corner zero modes. In this article, we review the elements of this novel second-order topological insulator, and demonstrate the essential physics by a simple model on the honeycomb lattice.
\end{abstract}


\maketitle

{\color{blue}\textit{Introduction}.} 
The symmetry-protected topological phases, such as topological (crystalline) insulators (TIs) and superconductors (TSCs),  have been one of the most active fields of physics during the last fifteen years~\cite{Volovik:book,Kane-RMP,XLQi-RMP,FuLiang2011prl,ShinseiRyu-RMP,Kruthoff2017prx,Benalcazar2017prb,LiuFeng2019prl,xie2021higher}. 
Based on the topological $K$ theory, the topological band theory has been established to classify and characterize various topological states \cite{Atiyah-KR,Kitaev2009AIP,Schnyder2008}. Symmetry plays an fundamental role in the classification of topological phases. Considering three discrete symmetries, namely time reversal $\mathcal{T}$, {charge conjugation} $\mathcal{C}$ and chiral symmetry $\mathcal{S}$, physical systems can be classified into ten symmetry classes, termed Altland-Zirnbauer (AZ) classes~\cite{AZ-Classification,Kitaev2009AIP,HoravaPRL05,Atiyah-KR,Schnyder2008,ZhaoYXWang14Septprb}, among which the eight ones with at least $\mathcal{T}$ or $\mathcal{C}$ are called real AZ classes. The topological classifications in the framework of the eight real AZ classes correspond to the real $K$ theory. Using the real $K$ theory, gapped systems including topological insulators and topological superconductors were first classified~\cite{Schnyder2008,Kitaev2009AIP,ShinseiRyu2010NJP}, and then gapless systems were classified as well~\cite{matsuura2013protected,ZhaoYXWang13prl,Chiu2014,Sato2014PRB}. All the classification tables exhibit an elegant eightfold periodicity along the dimensions for the eight real AZ classes.

After internal symmetries like $\mathcal{T}$ and $\mathcal{C}$, more and more spatial symmetries were involved to enrich symmetry-protected topological matter. It was noticed that combined symmetries $\P\T$ and $\C\P$ correspond to the orthogonal $K$ theory with $\P$ the spatial inversion, since they leave every $k$ point fixed in the reciprocal space. Hence, the topological classification table was worked out~\cite{ZhaoWang16Aprprl}. A remarkable feature is that groups $\Z$, $\Z_2$ and $0$ in the table appear in the reversed order in dimensionality, compared with previous tables for the real AZ classes. $\P\T$ and $\C\P$ are fundamental in nature, and therefore the classification table has been applied to explore topological phases in various physical systems, such as quantum materials \cite{ZhaoLu17Aprprl,Kruthoff2017prx,B-J-Yang19APRPRX}, topological superconductors~\cite{timm2017inflated,PhysRevLett.127.127001,tomonaga2021quasiparticle,lapp2020experimental}, and photonic/phononic crystals and electric-circuit arrays ~\cite{Mele2013prl,YangZJ2015prl,Ronny_2018np,Ozawa2019RMP,MaGC_2019nature,Serra_Garcia_2018nature,Yu_Zhao_NSR,Peterson_2018nature,lapp2020experimental}, and can generate unique topological structures with many novel consequences, such as non-Abelian topological charges, cross-order boundary transitions, and nodal-loop linking structures ~\cite{ZhaoLu17Aprprl,YuRui2015prl,B-J-Yang19APRPRX,ZhaoYang19prl,Wu1273,Wangzhijun2019prl,LiHN2020prl}. 

Remarkably, from the classification table, the symmetry class with $(\P\T)^2=1$ corresponds to the $\Z_2$ classification for $d=1$ and $d=2$. As revealed in Ref.\cite{ZhaoLu17Aprprl}, $(\P\T)^2=1$ leads to real band structures in contrast to conventional complex band structures. Then, the $\Z_2$ topological invariant $w_1$ for $d=1$ can be formulated as the quantized Berry phase in units of $\pi$ modulo $2\pi$. The case of $d=2$ is much fascinating. The topological invariant is the Euler number, a real version of the Chern number, for two valence bands. The Euler number is valued in $\Z$, but only its parity is stable if more trivial valence bands are added into consideration. The parity, namely the Euler number modulo $2$, is just the Stiefel-Whitney number $w_2$ in two dimensions, which determines whether the real vector bundle can be lifted into a spinor bundle.

The topological invariant $w_2$ gives rise to novel topological phases with extraordinary properties. In $3$D, it characterizes a real Dirac semimetal, which can be transformed into a nodal ring with symmetry-preserving perturbations. Then, the nodal ring is characterized by two topological charges $(w_1,w_2)$. In $2$D, it describes a topological insulator. The common topological wisdom is that the bulk topological invariant determines a unique form of the boundary modes, namely the well-known one-to-one bulk-boundary correspondence. However, a remarkably discovery in Ref.\cite{Wang2020} is that $w_2$ corresponds to multiple forms of boundary modes, extending the one-to-one correspondence to one-to-many. The $2$D topological insulator can host various second-order phases with odd $\P\T$-related pairs of corner zero-modes, which are mediated by first-order phases with helical edge states. Similarly, the $3$D semimetal can host second-order hinge Fermi arcs and first-order surface Dirac states as well. Recently, graphynes have been proposed as the material candidates which can realize both the $2$D topological insuslator and the $3$D topological semimetal~\cite{ZhaoYang19prl,chen2021graphyne,chen2022second}.

As aforementioned, the second-order phases of the $2$D topological insulator feature odd $\P\T$-related pairs of corner zero modes. It is interesting to look for its $3$D analog, which has been presented in Ref.\cite{Dai2020}.  Referring to the topological classification table for $\P\T$ and $\C\P$ symmetries, we notice that although the classification for $(\P\T)^2=1$ is trivial, with an additional chiral symmetry $\S$ with $\{\P\T,\S\}=0$ the classification is preserved as $\Z_2$ in $2$D and, more importantly, becomes nontrivial as $\Z_2$ in $3$D.  It is found that the corresponding topological invariants can be formulated in terms of Takagi's factorization. The topological invariant in $2$D is equivalent to $w_2$, while that in $3$D is a new topological invariant. Either in $2$D or in $3$D the bulk topological invariant can be manifested as odd $\P\T$-related pairs of corner zero-modes. Now, with the chiral symmetry, the two zero-modes in each pair are eigenstates with opposite eigenvalues of the chiral symmetry.

In this article, we review the elements of $2$D $\P\T$-protected topological insulators with or without chiral symmetry. The essential physics is demonstrated by the Honeycomb-lattice model, with only the nearest-neighbor hopping amplitudes. We show that under certain dimerization patterns the model is a topological insulator with nontrivial Stiefel-Whitney number or the Takagi topological invariant, and therefore presents all the nontrivial topological phenomena. Particularly, under various $\P\T$-invariant geometries, there are always odd $\P\T$-related pairs of corner zero-modes for the second-order topological phase.  Before diving into the details, it is noteworthy that the dimerized honeycomb model can be regarded as an abstraction from the graphynes~\cite{ZhaoYang19prl,chen2021graphyne,chen2022second}.

\textcolor{blue}{\textit{The honeycomb-lattice model.}} 
\begin{figure}[t]
	\includegraphics[scale=0.23]{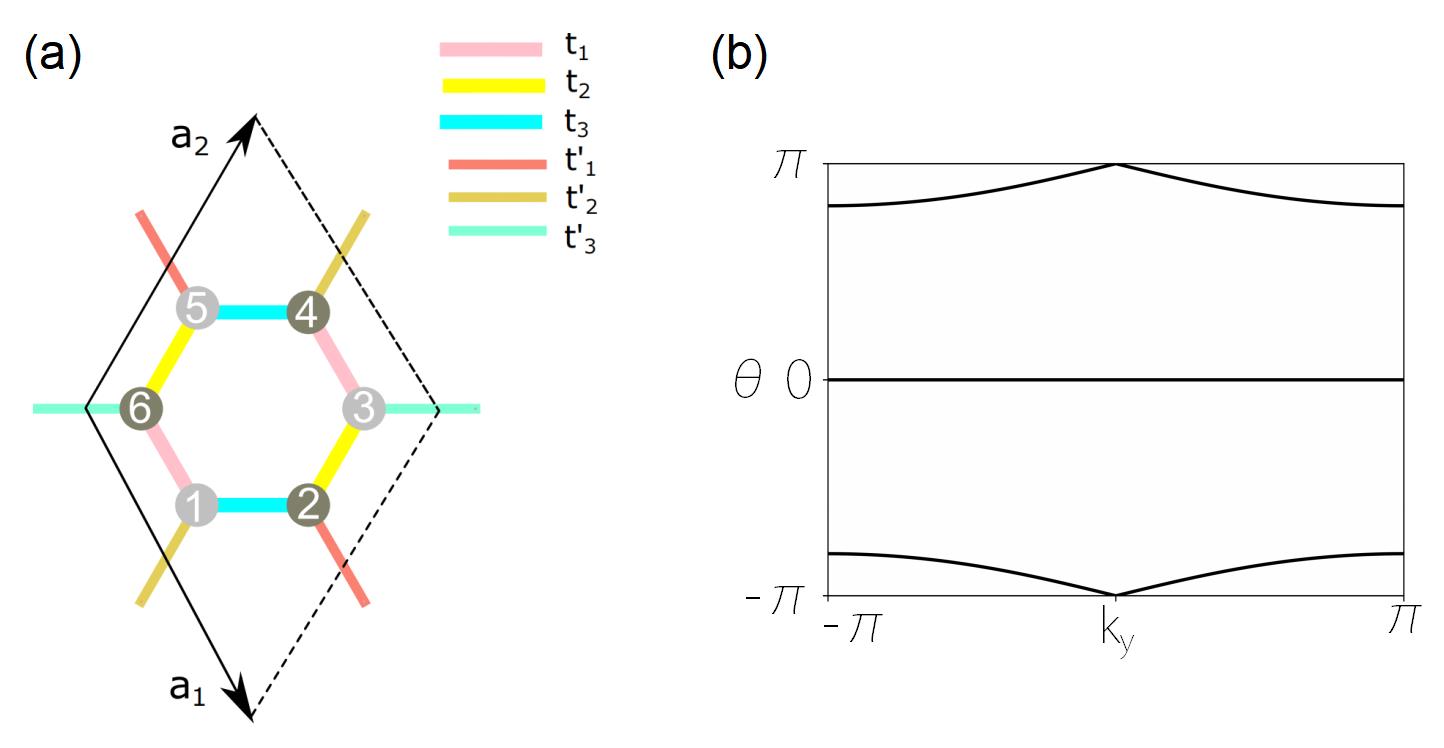}
	\caption{(a) Schematic of the honeycomb-lattice model. $t_i$ represents intracell hoppings and $t_i'$ represents intercell hoppings with $i=1,2,3$. The six atomic sites in a unit cell can be divided into two sublattice, as marked by the gray and brown circles, so that a site in one sublattice has all its nearest neighbors from  the other sublattice. (b) The winding of Wilson loop around $k_y$ for topological nontrivial case. The Wilson loop is computed along a large circle parametrized by $k_x$ for fixed $k_y$ (where $k_{x,y}$ represents periodic direction along $a_1,a_2$ or $a_3$). The loop exhibits a cross at $\theta=\pi$ and $k_y=0$, which means the $\Z_2$ topological invariant $\nu=1$. The parameters are set as  $t_1=t_2=t_3=1,t_1'=t_2'=t_3'=3$.}\label{lattice}
\end{figure}
Let us start with presenting the honeycomb-lattice model, the lattice structure is shown in Fig. \ref{lattice}(a). The Hamiltonian in momentum space is given by
\begin{equation}
\H(\k)=
\begin{bmatrix}
0&t_3&0&\chi^{(2)}_{\k} &0&t_1\\
t_3&0&t_2&0&\bar{\chi}^{(1)}_{\k} &0\\
0&t_2&0&t_1&0&\chi^{(3)}_{\k}\\
\bar{\chi}^{(2)}_{\k}&0&t_1&0&t_3&0\\
0&\chi^{(1)}_{\k}&0&t_3&0&t_2\\
t_1&0&\bar{\chi}^{(3)}_{\k}&0&t_2&0
\end{bmatrix},
\label{Hamiltonian}
\end{equation} 
where $\chi^{(i)}_{\k}=t_i^\prime e^{-i\k\cdot \bm{a}_i}$ with $i=1,2,3$.  Here, $\bm{a}_i$ are the bond vectors connecting the centers of nearest-neighbor unit cells, as indicated in Fig.\ref{lattice}(a) with $\sum_i \bm{a}_i=0$.
The Hamiltonian has inversion symmetry with $\PP=\sigma_1\otimes I_3\I$, spinless time-reversal symmetry with $\TT=\K\I$, and therefore spacetime-inversion symmetry with $\PP\TT=\sigma_1\otimes I_3\K$, where $\K$ is the complex conjugation and $\I$ is the inversion of momenta. Note that $\sigma$'s are the Pauli matrices acting on the sublattice space [see Fig.\ref{lattice}(a)], and $I_3$ is the $3\times 3$ identity matrix. Each inversion center is taken as the center of a hexagon in real space. The sublattice symmetry operator is $\hat{\S}=I_3\otimes\sigma_3$. Since the inversion exchanges sublattices, both $\P$ and $\P\T$ anti-commute with $\S$, namely, $\{\hat{\P},\hat{\S}\}=\{\PP\TT,\hat{\S}\}=0$. 

To obtain the nontrivial topological phases, we calculate the determinant of the Hamiltonian \eqref{Hamiltonian} at $\Gamma$ point \cite{Gamma-Point} in the Brillouin zone as
\begin{equation}
\mathrm{det}[\H(\Gamma)]=-(t_1^2t_1'+t_2^2t_2'+t_3^2t_3'-2t_1t_2t_3-t_1't_2't_3')^2.
\end{equation}
Since the bulk topological criticality generally corresponds gap-closing point, we can obtain the topological phase-transition points by letting $\mathrm{det}[\H(\Gamma)]=0$, which gives
\begin{equation}
t_1^2t'_1+t_2^2t'_2+t_3^2t'_3=2t_1t_2t_3+t'_1t'_2t'_3.\label{bulk-criticality}
\end{equation}
Interestingly, if \eqref{bulk-criticality} holds, the system is generally reduced to a topologically equivalent graphene model with two Dirac points in the first Brillouin zone \cite{Haldane1988prl}.
When $t_1^2t'_1+t_2^2t'_2+t_3^2t'_3<2t_1t_2t_3+t'_1t'_2t'_3$, the system steps into a topological phase, while conversely the system becomes a trivial phase, which can be checked by computing Stiefel-Whitney number or Takagi's factorization. 

{\color{blue}\textit{Topological invariants}} The topology can be determined by various formulas of the topological invariant. We now briefly review them. First, as given in Ref.\cite{ZhaoLu17Aprprl}, the topological invariant can be determined by the Wilson loop
\begin{equation}
\W(k_y)=P\exp \bigg(-i\int_{C_{k_y}}dk_x~\mathcal{A}(k_x,k_y)\bigg)
\end{equation}
(with $P$ indicating the path order) along large circles parametrized by $k_x$.
$C_{k_y}$ is the contour at a fixed $k_y$ and $\mathcal{A}(k_x,k_y)$ is the non-Abelian Berry connection for the valence bands. The topological information is encoded in the phase factors $\theta(k_y) \in\left(-\pi, \pi \right] $ of the $N$ eigenvalues $\lambda_m(k_y)$ of $\W(k_y)$ for valence bands:
\begin{equation}
\theta_m(k_y)=\Im[\log\lambda_m(k_y)].\label{theta}
\end{equation}
Different from the conventional TIs and Chern insulators, the Wilson loop spectral flow for real phases are mirror symmetric with respect to the $\theta=0$ axis [see Fig. \ref{lattice}(b)]. This is because $\W(k_y)$ is equivalent to a mapping from $k_y\in S^1$ to $O(N)$ up to a unitary transformation~\cite{ZhaoLu17Aprprl}. The topological information can be pictorially derived from counting how many times $\zeta$ the trajectories cross $\theta=\pi$ as
\begin{equation}
	w_2=\zeta \mod 2.
\end{equation}
For honeycomb lattice with $t_1^2t'_1+t_2^2t'_2+t_3^2t'_3<2t_1t_2t_3+t'_1t'_2t'_3$,  a single crossing exsit as shown in Fig.\ref{lattice}(b), namely, $w_2=1$, which indicates the model is in a topological nontrivial phase.

As aforementioned, our system is protected by spacetime inversion symmetry $\P\T$ and sublattice (chiral) symmetry $\S$. These symmetries constraint the classifying space of $\H(\k)$ to be symmeric unitary matrices. Thus the $\Z_2$ invariant from the Takagi's factorization can be defined~\cite{Dai2020}, which leads to an alternative formulation for $w_2$. We now prove the equivalence of the two formulas. For technical simplicity, we assume the momentum space as a sphere $S^2$, which is sufficient to present the essential ideas.

\begin{figure}[t]
	\includegraphics[scale=0.26]{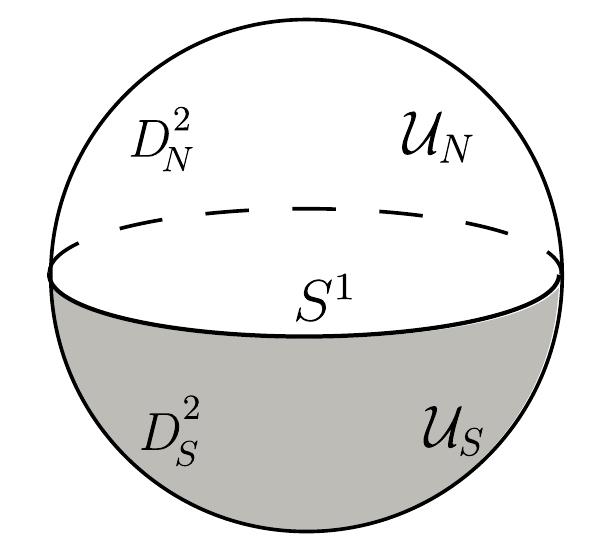}
	\caption{The Takagi factors in $2$D.}
	\label{CI2}
\end{figure}

In general, $\S$ requires the Hamiltonian $\H(\k)$ to be block anti-diagonal and $\P\T$ requires the upper-right block to be symmetric. Thus the flattened Hamiltonian $\tilde\H(\k)$ is given by
\begin{equation}
\tilde\H(\k)=\begin{bmatrix}
0 &\Q(\k) \\ \Q^\dagger(\k) & 0
\end{bmatrix},~\Q=\Q^T,~\Q\Q^\dagger=I_M,\label{Takagi_2D}
\end{equation}
where $\Q(\k)=\UU(\k)\UU^T(\k)$ is a unitary symmetric matrix for each $\k$ and $M$ denotes the number of valence (conduction) bands. $\UU(\k)\in U(M)$ is the Takagi factor. The classifying space for this symmetric class is ${US}(M)=U(M)/O(M)$ \cite{Dai2020}.
Here, $\pi_2[{US}(M)]=\Z_2$ corresponds to the topological invariant of our system. Consider a $2$D sphere $S^2$, which is divided into north and south hemispheres $D^2_{N,S}$, overlapping along the equator $S^1$. The Takagi factors $\UU_{N/S}$ over $D_{N/S}^2$, respectively, can be transformed to each other by a gauge transformation $\mathcal O_{S^1}$ over the equator $S^1$, as shown in Fig.~\ref{CI2}. 
$\mathcal O_{S^1}$ is given by $$\mathcal O_{S^1}=\UU^\dagger_N|_{S^1}\UU_S|_{S^1}, ~\mathcal O_{S^1}\in O(M).$$ $\pi_1[{O}(M)]=\Z_2$ for $M>2$ leads to obstructions for a global Takagi's factorization over $S^2$.

The conduction and valence wavefunctions of $\tilde\H(\k)$ can be given by
\begin{equation}
|+,n\rangle=\frac{1}{\sqrt 2}\begin{bmatrix}
\UU\varphi_n\\\ \UU^*\varphi_n
\end{bmatrix},~
|-,n\rangle=\frac{i}{\sqrt 2}\begin{bmatrix}
\UU\varphi_n\\\ -\UU^*\varphi_n
\end{bmatrix},
\end{equation}
where $n\in\{1,2,\cdots,M\}$. $\varphi_n=(0~0~\cdots~0~1~0~0~\cdots~0)^T$ is a unit vector with ``$1$'' locating at the $n$-th position.

Performing a unitary transformation $\UU_R=e^{-i{\pi}/{4}}e^{i{\pi\sigma_1}/{4}}$ on this system, the Hamiltonian and valence wavefunctions both become real. Meanwhile, $\P\T$ and $\S$ are transformed to $\K$ and $\sigma_2$, respectively.
Over the intersection $S^1$, transition function $t_{S^1}$  of real valence wavefunctions can be given by
\begin{equation}
[t_{S^1}]_{mn}=\langle-,m|_N\big{|}_{S^1}\UU_R^\dagger  \UU_R|-,n\rangle_S\big{|}_{S^1}=\mathcal [\mathcal O_{S^1}]_{mn}.
\end{equation}
Thus, we know the transition function $t_{S^1}$ of real valence wavefunctions is equal to the gauge transformation $\mathcal O_{S^1}$. As noted in Ref.\cite{ZhaoLu17Aprprl}, $w_2$ is just the parity of the winding number of the transition function for valence bands. Thus, we see the equivalence of two $2$D topological invariants.

{\color{blue}{\textit{Physical consequence}.}} According to analytical and numerical methods, we reveal that three pairs of hopping parameters $t_i$ and $t'_i$ (with $i=1,2,3$) jointly determine the configuration of topological boundary modes. To facilitate  understanding the relation between distinct boundary modes and parameters, we define a boundary effective mass term $m_i$ for each edge:
\begin{equation}
	m_i=t_it'_i-t_jt_k \quad \text{with} \quad i\neq j\neq k,      \label{effective-mass}
\end{equation}
where the subscript $i$ denotes the hopping along the primitive vector $\mathbf{a}_i$ direction ($\mathbf{a}_3=-\mathbf{a}_1-\mathbf{a}_2$). The above Eq.\eqref{effective-mass} can be derived from the boundary effective Hamiltonian \cite{Edge-Mass}.
Hence, if $m_i=0$, the corresponding edges are gapless, which is also the boundary critical condition to separate two second-order topological phases.\par 

\begin{figure}[t]
	\includegraphics[scale=0.21]{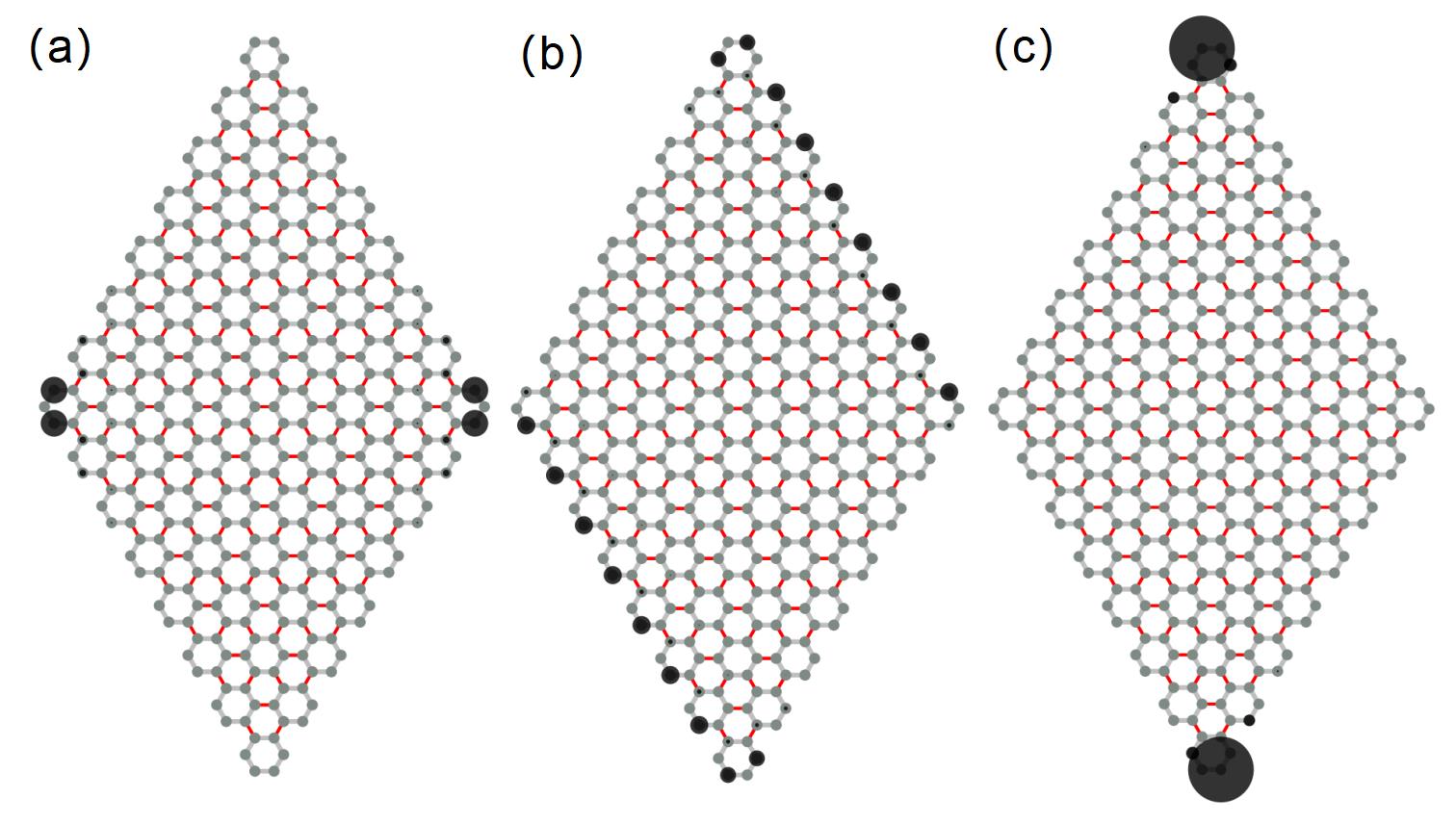}
	\caption{(a)-(c) Possible topological boundary modes for the rhombic-shaped sample with $10\times10$ unit cells. Black circles indicate the distribution of density of zero-mode (a)(c) Second-order TI phases with a single pair of zero-modes corners in diagonal and off-diagonal (or horizontal and vertical) directions, respectively. (b) Helical edge modes with the boundary effective mass $m_i=0$, which is a critical state separating two second-order TI phases. Parameters are set as (a) $t_{1,2,3}=1,t'_{1,2,3}=3$, (b) $t_1=1,t_2=2,t_3=1.5,t'_{1,2,3}=3$,  (c) $t_1=1.8,t_2=0.2,t_3=0.8,t'_{1,2,3}=3$.}\label{rhombic}
\end{figure} 

To demonstate the boundary modes, we consider a rhombic-shaped $2$D sample with armchair termination, i.e., by opening boundary along $\mathbf{a}_1$ and $\mathbf{a}_2$ direction, as shown in Fig.~\ref{rhombic}. If $m_1=0$ and $m_2\neq0$, the helical edge modes along periodic $\mathbf{a}_1$ can be obtained, as shown in Fig.~\ref{rhombic}(b). However, once $m_{1,2}\neq 0$, the helical edge modes will be gapped and the localized corner modes will emerge. More specifically, for the case with $\mathrm{sgn}(m_1)=\mathrm{sgn}(m_2)$ ($\mathrm{sgn}(m_1)=-\mathrm{sgn}(m_2)$), the corner modes will locate at $120^\circ$ ($60^\circ$) corners, as shown in Fig.\ref{rhombic}(a) and (c) respectively. The $\P\T$ symmetry requires that the corner zero-modes always come in pairs and the chiral symmetry sets the midgap modes exactly at zero energy \cite{Finite-size}.

To keep the completeness of honeycomb unit cell in a rhomboid sample, one only has three kinds of armchair edges, namely the edge parallel to  $\mathbf{a}_i$ direction with $i=1,2,3$. If the edge connected by the same corner has the same mass term $m_i$ sign, the corner zero-modes will be localized at the obtuse angle of the rhomboid, otherwise at acute corners. We shall theoretically explain these numerical results in the next section. It is emphasized that in the whole process of the edge-phase transitions, the bulk gap is always open and the symmetries are preserved, therefore, the bulk invariant $\nu$ is unchanged. Thus the conventional bulk-boundary correspondence is not appliable for TTI, namely, the bulk invariant can not uniquely determine the boundary modes, but dictates an edge criticality, as the concept  mentioned in previous work \cite{Wang2020}.

\begin{figure}[t]
	\includegraphics[scale=0.19]{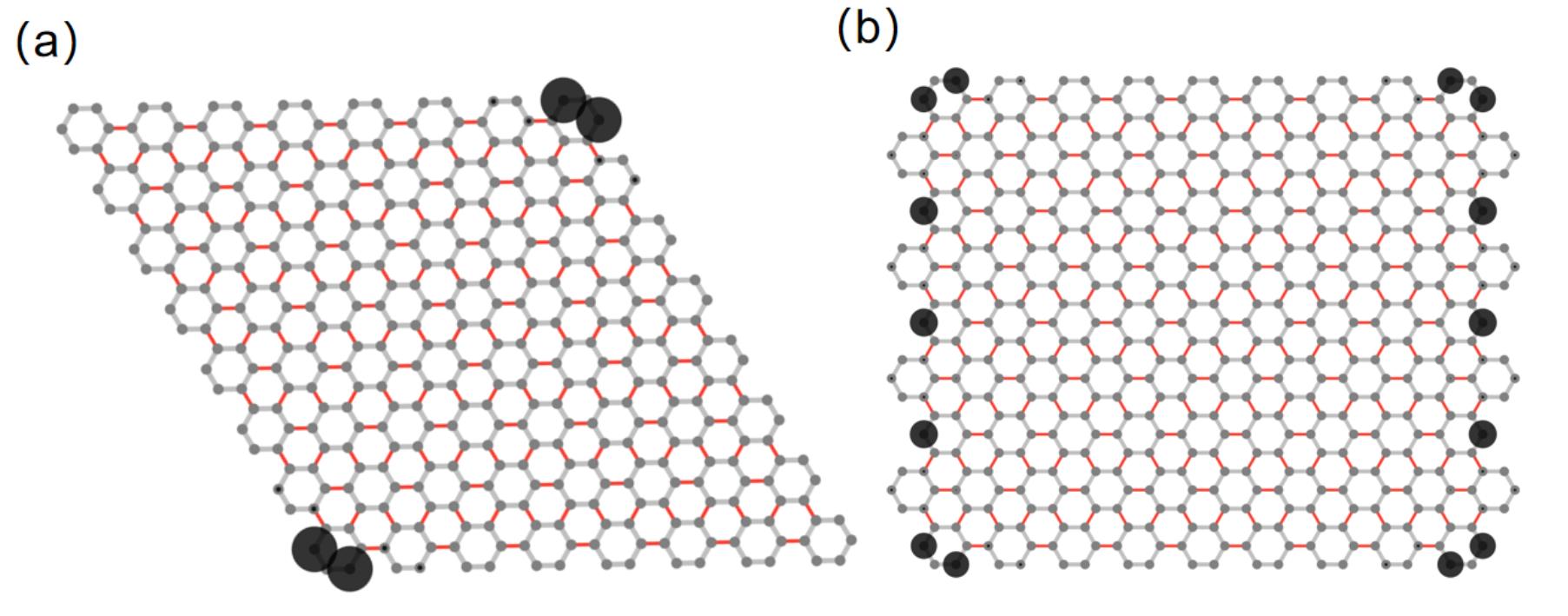}
	\caption{(a) Topological corner modes in rhombic sample with $t_i=1,t'_i=3$. (b) Topological edge modes in rectangle sample with same parameters as rhombic($t_{1,2,3}=1,t'_{1,2,3}=3$).}\label{boundary}
\end{figure}

\begin{figure}[b]
	\includegraphics[scale=0.225]{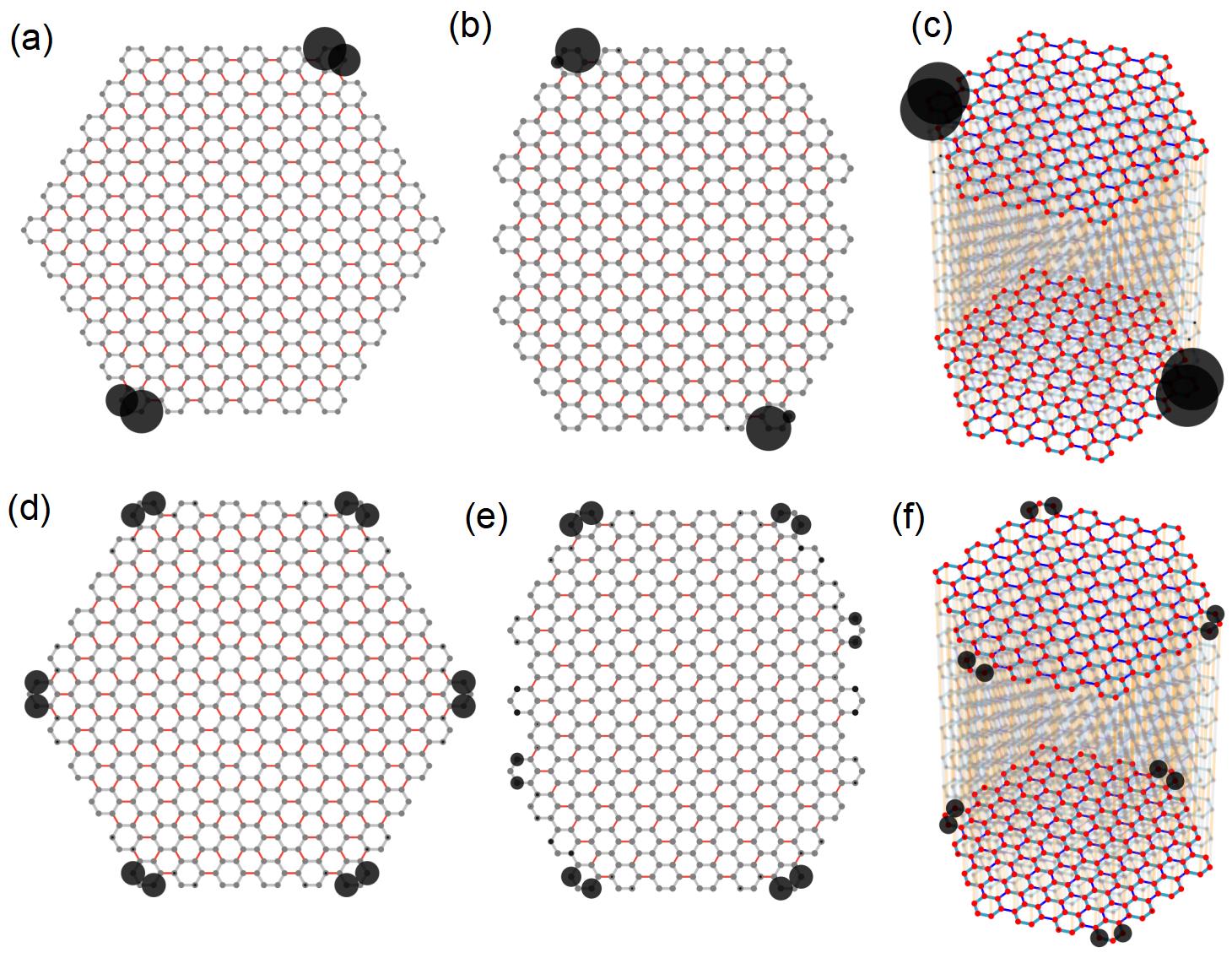}
	\caption{(a),(d) Orthohexagonal sample with one and three pairs of zero-mode corners respectively. (b),(e) Octagonal sample with one and three pairs of zero-mode corners respectively. (c),(f)hexagonal prisms sample with one and three pairs of corner zero-modes respectively.  }	\label{odd-pairs}
\end{figure}

As promised in \emph{introduction}, we now proceed to tune the boundary modes with fixed parameters. 
In the rhombic case, all samples terminate with armchair edges and exhibit parameter-depended boundary modes. As long as $\P\T$ and $\S$ are not violated, the finite samples can be cut with not only rhomb as shown in Fig.~\ref{rhombic}, but also hexagon(see Fig.\ref{odd-pairs} ). Beside armchair edges, zigzag edges can serve as termination too. Creatively, with fixed parameters but different boundary selections, one can also find various distinguishable boundary modes. For instance, helical edge states emerge on the zigzag edges in a rectangle sample as shown in Fig. \ref{boundary}(b), with the same hopping parameters as Fig. \ref{boundary}(a). This result further proves that the bulk topological invariant can not uniquely determine the topological boundary modes. We also study lots of other patterns with the same parameters, and abundant topological boundary modes consisting of corner zero modes and gapless edge modes can be obtained (see Appendix.~\ref{different_boundary}). They are all boundary-selection-depended. Hence, we propose that one can obtain needed topological boundary modes by choosing particular boundary geometry, without tuning parameters, which is usually difficult to perform in real systems.   

Novelly, in both situations discussed above, for second-order topological phases, the number of the zero-energy corners must be odd pairs. For example, for a hexagonal sample, we can only find one or three pairs of zero-mode corners, as shown in Fig. \ref{odd-pairs}(a) and (b). Similarly, the Octagonal sample also has one or three pairs of zero-mode corners as shown in Fig. \ref{odd-pairs}(c) and (d). 


We find that the peculiarity of odd-pair-zero-modes is universal and it can be generalized to a higher-dimensional situation, such as 3D~\cite{Dai2020}. The 3D model is constructed by stacking the 2D honeycomb TTI discussed above in a staggered manner to preserve the sublattice symmetry $\S$. The details of the construction of the model can be found in Appendix.~\ref{3dmodel} .  The inversion center is chosen as the center of hexagonal in one layer. 
Thus the anti-commuting relation of $\P\T$ and $\S$ are preserved. We cut a finite hexagonal prisms sample that keeps the symmetries. One can find only odd pairs (one or three) of corner states related by $\P\T$ appear. Note that the corner zero-modes can be driven to other corners by tunning the hopping parameters like in a 2D situation.

{\color{blue} \textit{Analytic method.}} We first proceed to solve the boundary criticality along the periodic $\mathbf{a}_1$ direction and openning boundary with $\mathbf{a}_2$. Replace $e^{-i\k\cdot \mathbf{a_2}}$ by $S$ in Hamiltonian \eqref{Hamiltonian}, with $S$ ladder operator and $S|i\rangle=|i+1\rangle,S^\dagger|i\rangle=|i-1\rangle$. Then $e^{-i\k\cdot\mathbf{a}_3}$ can be represented by $e^{i\k\cdot\mathbf{a}_1}S^\dagger$ since $\mathbf{a}_3=-(\a_1+\a_2)$. After a series of tedious derivation (see Appendix.~\ref{A}), we obtian the effective Hamiltonian of the bottom boundary:
\begin{equation}
	\begin{split}
		\H_B(\k\cdot \a_1)=\begin{bmatrix}
			0&t_3&0&t_1\\
			t_3&0&t_1'e^{i\k\cdot \mathbf{a}_1}&0\\
			0&t_1'e^{-i\k\cdot \mathbf{a}_1}&0&t_2\\
			t_1&0&t_2&0
		\end{bmatrix}.
	\end{split}\label{bottom-H}
\end{equation}
Following the same argument with aforementioned bulk criticality, we can obtain the boundary criticality by letting $\mathrm{det}[\H_B(\k\cdot \a_1)]=0$, which leads to
\begin{equation}
	t_1t_1'-t_2t_3=0.\label{effective-mass-a1}
\end{equation}
The above Eq. \eqref{effective-mass-a1} holds only at $\k\cdot\a_1=0$.
Thus, when the system is in a topological nontrivial case, Eq. \eqref{effective-mass-a1} related edge criticality separates two different second-order topological phases with corner zero modes. 
Likely, we can obtain similar results for periodic $\a_2$ and $\a_3$ directions. 
For convenience, we can define boundary effective mass by the left of Eq. \eqref{effective-mass-a1} for edges parallel to $\a_1$. Or, generally Eq. \eqref{effective-mass} for edges parallel to $\a_i$. Different from the armchair edges, the zigzag terminations has an additional boundary criticality, namely, the effective masses can be defined by
\begin{equation}
	M_i=\frac{1}{2}\sum_{j,k}\epsilon_{ijk}(t_j^2t'_j-t_k^2t'_k).
\end{equation}
The derived details can be found in the Appendix.~\ref{square}.

With the effective mass orderly distributing on each edge \cite{Edge-Mass}, the existence of corner zero-modes is reduced to a Jackiw-Rebbi problem \cite{Jackiw1976prd}. The corners with opposite effective masses on both sides can have zero-modes.

{\color{blue} \textit{Discussion.}} In this article, we present a simple 2D realizable honeycomb-lattice model to demonstrate the essential physics of the Takagi topological insulator. It is found that with unchanged topological invariant, one can tune topological boundary modes by not only parameters, but also boundary selections. It goes beyond the common wisdom about bulk-boundary correspondence, and gives rise to much richer boundary physics. 

Our model with novel physics is closely related to real systems. It is easier to realize our model by photonic/phononic crystals, electric-circuit arrays and mechanics systems, since only have nearest-neighbor hopping amplitudes are included into the model. Several special cases of our model have been recently realized in photonic/phononic crystals \cite{PhysRevLett.125.255502,photonic}, where hopefully the general form of our model can be further experimentally examined.

\begin{acknowledgements}
	The authors acknowledge the support from the National Natural Science Foundation of China under Grants (No.11874201, No.12174181, and No.12161160315).
\end{acknowledgements}

\appendix

\section{Theoretical method for critical conditon of  rhombic sample }\label{A}

The lattice  is shown as FIG.1 in main text. The Hamiltonian $\H$ in $2$D momentum space is given in Eq.~\eqref{Hamiltonian}. The determinant of $\H$ is given as
\begin{equation*}
\begin{aligned}
		&\det(\H)=-EE^*,\\
		&E=-2t_1t_2t_3-t'_1t'_2t'_3+t_1^2\bar{\chi}^{(1)}_{\k}+t_2^2\chi^{(2)}_{\k}+t_3^2\chi^{(3)}_{\k}. 
\end{aligned}
\end{equation*}
For simplicity, we write $\k\cdot\mathbf{a}_i$ as $k_i$. To make $\H$ gapless, we obtain
\begin{eqnarray}
	\begin{split}
		&2t_1t_2t_3+t'_1t'_2t'_3\\
		&~~~~~-(t_1^2t'_1\cos k_1+t_2^2t'_2\cos k_2+t_3^2t'_3\cos k_3)=0,\\
		&t_1^2t'_1\sin k_1-t_2^2t'_2\sin k_2-t_3^2t'_3\sin k_3=0.
		\end{split}
	\label{Condition}
\end{eqnarray}
Because the hopping terms are real and positive, we know that
\begin{equation}
	\begin{split}
		&\min(2t_1t_2t_3+t'_1t'_2t'_3\\
		&-(t_1^2t'_1\cos k_1+t_2^2t'_2\cos k_2+t_3^2t'_3\cos k_3))\\
		=&2t_1t_2t_3+t'_1t'_2t'_3-(t_1^2t'_1+t_2^2t'_2+t_3^2t'_3),
	\end{split}
\end{equation}
with $k_1=k_2=k_3=0$. Meanwhile, the second formula in the Eq.~\eqref{Condition} always holds. Therefore, $2t_1t_2t_3+t'_1t'_2t'_3-(t_1^2t'_1+t_2^2t'_2+t_3^2t'_3)=0$ corresponds to the gapless phase, which is also the critical phase between trival and nontrival  phases.
\begin{figure}[h]
	\includegraphics[scale=0.2]{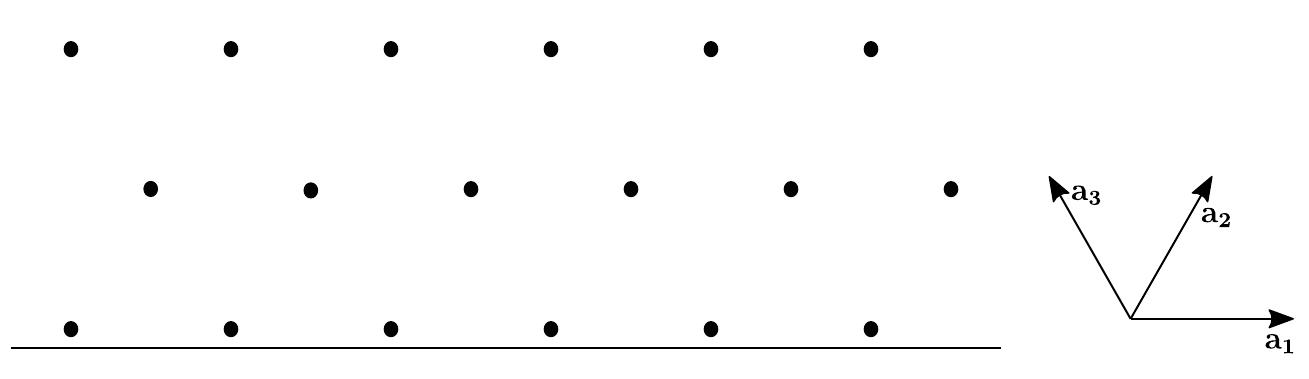}
	\caption{Each point represents a unit cell. The lattice is semi-infinite in the direction perpendicular to the vector $\mathbf{a}_1$.}
	\label{Boundary}
\end{figure}
Opening the boundary parallel to the lattice vector $\mathbf{a}_1$ as FIG.~\ref{Boundary}, the Hamiltonian is transformed to 
\begin{equation*}
	\H_2=\begin{bmatrix}
		0 & t_1 & 0 & t'_2S_2^\dagger & 0 & t_3\\
		t_1 & 0 & t_2 & 0 & t'_3e^{ik_1}S_2^\dagger & 0\\
		0 & t_2 & 0 & t_3 & 0 & t'_1e^{ik_1}\\
		t'_2S_2 & 0 & t_3 & 0 & t_1 & 0\\
		0 & t'_3e^{-ik_1}S_2& 0 & t_1 & 0 & t_2\\
		t_3 & 0 & t'_1e^{-ik_1} & 0 & t_2 & 0
	\end{bmatrix},
\end{equation*}
where $S_2$ and $S_2^\dagger$ are the forward and backward translation operators along the $\mathbf{a}_2$-direction, respectively. The actions of $S_2$ and $S_2^\dagger$ on the real space basis of the tight-binding model for the the $\mathbf{a}_2$-direction are given by
\begin{equation}
	S_2|i\rangle=|i+1\rangle,\quad S_2^\dagger|i\rangle=|i-1\rangle,\quad S_2^\dagger|0\rangle=0,
\end{equation}
where the nonnegative integer i labeling the lattice site along the $\mathbf{a}_2$-direction.  Accordingly, the matrices can be explicitly written as
\begin{equation}
	S_2=\begin{bmatrix}
		0&0&0&0&\cdots\\
		1&0&0&0&\cdots\\
		0&1&0&0&\cdots\\
		0&0&1&0&\cdots\\
		\vdots&\vdots&\vdots&\vdots&\ddots
	\end{bmatrix},\quad 
	S_2^\dagger=\begin{bmatrix}
		0&1&0&0&\cdots\\
		0&0&1&0&\cdots\\
		0&0&0&1&\cdots\\
		0&0&0&0&\cdots\\
		\vdots&\vdots&\vdots&\vdots&\ddots
	\end{bmatrix}.\nonumber
\end{equation}

Adopting the \emph{Ans$\ddot{a}$tze}
\begin{equation*}
	|\psi(k_1)\rangle=\sum_{i=0}^\infty \lambda^i|i\rangle\otimes|\xi(k_1)\rangle,\quad\H_2|\psi(k_1)\rangle=\mathcal{E}|\psi(k_1)\rangle
\end{equation*}
with $|\lambda|<1$ for the boundary states. In the bulk with $i\ge 1$, we obtain that
\begin{widetext}
\begin{equation}
	\begin{bmatrix}
		0 & t_1 & 0 & t'_2\lambda & 0 & t_3\\
		t_1 & 0 & t_2 & 0 & t'_3e^{ik_1}\lambda & 0\\
		0 & t_2 & 0 & t_3 & 0 & t'_1e^{ik_1}\\
		t'_2\lambda^{-1} & 0 & t_3 & 0 & t_1 & 0\\
		0 & t'_3e^{-ik_1}\lambda^{-1}& 0 & t_1 & 0 & t_2\\
		t_3 & 0 & t'_1e^{-ik_1} & 0 & t_2 & 0
	\end{bmatrix}\begin{bmatrix}
		\xi_1\\\xi_2\\ \xi_3\\\xi_4\\\xi_5\\\xi_6
	\end{bmatrix}=\mathcal E\begin{bmatrix}
		\xi_1\\\xi_2\\ \xi_3\\\xi_4\\\xi_5\\\xi_6
	\end{bmatrix}.\label{I1}
\end{equation}
\end{widetext}

Restricting to the surface layer with $i = 0$, we obtain that
\begin{equation}
	\begin{bmatrix}
		0 & t_1 & 0 & t'_2\lambda & 0 & t_3\\
		t_1 & 0 & t_2 & 0 & t'_3e^{ik_1}\lambda & 0\\
		0 & t_2 & 0 & t_3 & 0 & t'_1e^{ik_1}\\
		0 & 0 & t_3 & 0 & t_1 & 0\\
		0 & 0& 0 & t_1 & 0 & t_2\\
		t_3 & 0 & t'_1e^{-ik_1} & 0 & t_2 & 0
	\end{bmatrix}\begin{bmatrix}
		\xi_1\\\xi_2\\ \xi_3\\\xi_4\\\xi_5\\\xi_6
	\end{bmatrix}=\mathcal E\begin{bmatrix}
		\xi_1\\\xi_2\\ \xi_3\\\xi_4\\\xi_5\\\xi_6
	\end{bmatrix}.\label{I2}
\end{equation}
The difference of Eqs. \eqref{I1} and \eqref{I2} gives $\xi_1=\xi_2=0$. Thus, the effective Hamiltonian for the boundary state is
\begin{equation}
	\H_{eff}^1=\begin{bmatrix}
		0 & t_3 & 0 & t'_1e^{ik}\\
		t_3 & 0 & t_1 & 0\\
		0 & t_1 & 0 & t_2\\
		t'_1e^{-ik} & 0 & t_2 & 0
	\end{bmatrix}.
\end{equation}
The determinant of $\H_{eff}^1$ is given by
\begin{equation}
	\det(\H_{eff}^1)=t_2^2t_3^2+t_1^2(t'_1)^2-2t_1t_2t_3t'_1\cos k.
\end{equation}
Therefore, we know $\det\H_{eff}^1=0$ with $k_1=0$ only when $t_2t_3=t_1t'_1$. In other words, the boundary effective Hamiltonian $\H_{eff}^1$ corresponds to a gapless phase only if $t_2t_3=t_1t'_1$. By the same way, we obtain the gapless boundary along the direction of $\mathbf a_2$ and $\mathbf a_3$ with $t_1t_3=t_2t'_2$ and $t_1t_2=t_3t'_3$, respectively.\\

\section{Theoretical method for critical conditon of  square-shaped sample}\label{square}

Here, we present the  square-shaped and parallelogram-shaped boundary conditions as examples to derive the critical points for parameters.

\begin{figure}
	\includegraphics[scale=0.236]{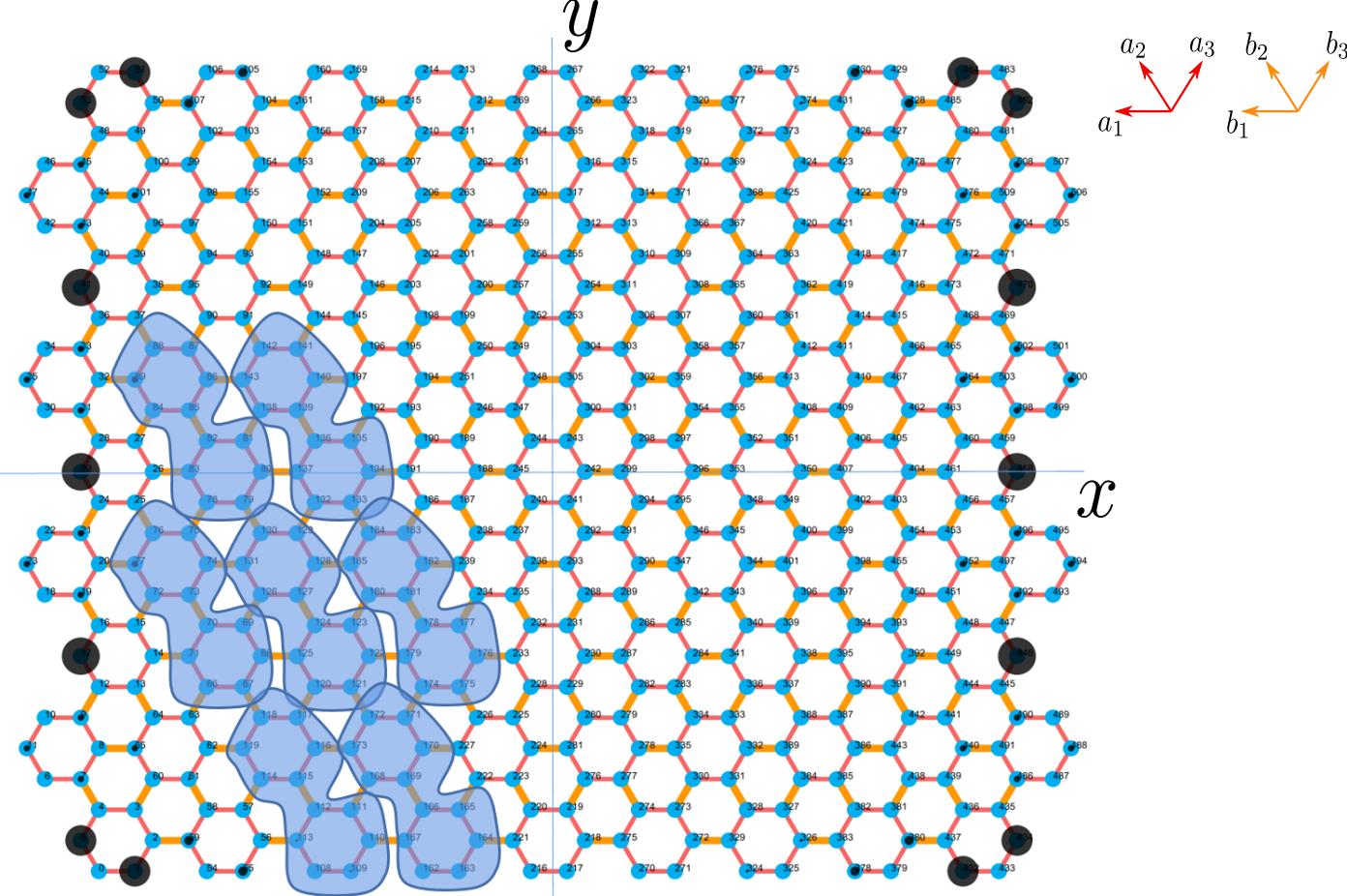}
	\caption{The square-shaped lattice. For translational invariance along the horizontal and vertical directions, we take 12 sites as an  unit cell.}
\end{figure}

The Hamiltonian $\H(\k)$ in reciprocal space can be written as
\begin{widetext}
\begin{equation}
	\begin{bmatrix}\nonumber
		0&a_1&0&0&0&a_2&0&0&b_3e^{-ik_y}&0&0&0\\
		a_1&0&a_3&0&0&0&0&0&0&b_2e^{i(k_x-k_y)}&0&0\\
		0&a_3&0&a_2&0&b_1e^{ik_x}&0&0&0&0&0&0\\
		0&0&a_2&0&a_1&0&0&0&0&0&0&b_3e^{ik_x}\\
		0&0&0&a_1&0&a_3&b_2&0&0&0&0&0\\
		a_2&0&b_1e^{-ik_x}&0&a_3&0&0&0&0&0&0&0\\
		0&0&0&0&b_2&0&0&a_3&0&0&0&a_1\\
		0&0&0&0&0&0&a_3&0&a_2&0&b_1e^{ik_x}&0\\
		b_3e^{ik_y}&0&0&0&0&0&0&a_2&0&a_1&0&0\\
		0&b_2e^{-i(k_x-k_y)}&0&0&0&0&0&0&a_1&0&a_3&0\\
		0&0&0&0&0&0&0&b_1e^{-ik_x}&0&a_3&0&a_2\\
		0&0&0&b_3e^{-ik_x}&0&0&a_1&0&0&0&a_2&0
	\end{bmatrix}
\end{equation}
\end{widetext}
To find the critical condition for topological phase transition, we calculate the determinant of the bulk Hamiltonian, 
\begin{equation*}
	\begin{split}
		&\mathrm{det}(\H(\k))=-(2t_1t_2t_3+t'_1t'_2t'_3-(t_1^2t'_1\cos k_x\\&
		+t_2^2t'_2\cos k_y+t_3^2t'_3\cos(k_x-k_y))^2\\
		&-(t_1^2t'_1\sin k_x-t_2^2t'_2\sin k_2-t_3^2t'_3\sin(k_x-k_y))^2\le 0.\\
	\end{split}
\end{equation*}
So, if the Hamiltonian has zero energy, det$(\H(\k))$  must be zero and 
\begin{equation}
	\begin{split}
		&A=2t_1t_2t_3+t'_1t'_2t'_3-(t_1^2t'_1\cos k_x+t_2^2t'_2\cos k_y\\&
		~~~~~~~~~~+t_3^2t'_3\cos(k_x-k_y))=0,\\
		&B=t_1^2t'_1\sin k_x-t_2^2t'_2\sin k_2\\&~~~~~~~~~~-t_3^2t'_3\sin(k_x-k_y)=0.\label{AB=0}
	\end{split}
\end{equation}
On the other hand, 
\begin{equation}
	\begin{split}
		&A=2t_1t_2t_3+t'_1t'_2t'_3-(t_1^2t'_1\cos k_x+t_2^2t'_2\cos k_y\\ &+t_3^2t'_3\cos(k_x-k_y))\\
		&\le2t_1t_2t_3+t'_1t'_2t'_3-(t_1^2t'_1+t_2^2t'_2+t_3^2t'_3),
	\end{split}
\end{equation}
if and only if $k_x=k_y=0(k_{x,y}\in [-\pi,\pi])$, the equal sign establishes in the inequality. And at this moment, $B=0$ also holds. 
Hence the condition of det$(\H_{\k})=0$ is 
\begin{equation}
	t_1^2t'_1+t_2^2t'_2+t_3^2t'_3=2t_1t_2t_3+t'_1t'_2t'_3.\label{bulk-condition}
\end{equation}
When the above equation holds, there exist gapless bulk states, more explicitly, there is a Dirac point localized at $k_x=k_y=0$. So this condition may be the critical point between trivial and nontrivial topological phases. We check the parameter condition \eqref{bulk-condition} with numerical wilson loop method and find the condition \eqref{bulk-condition} is the critical point exactly.

{\color{black} Note that although the Eqs. \eqref{AB=0} has solutions beyond $\Gamma$ point, such as $(\pi,0),(0,\pi),(\pi,\pi)$, we check that the parameters conditions for these zero-energy points do not distingush the trivial and non-trivial topological phases.}

\subsection{The zigzag boundary}

We first study zigzag boundary, namely, $y$ direction is periodic. In this case, by replacing $e^{-ik_x}$ with $S$, where $S$ is ladder operator and $S|i\rangle=|i+1\rangle,S^\dagger|i\rangle=|i-1\rangle,S^\dagger|0\rangle=0$. 
More explicitly, the semi-infinite translation operators are now written as
\begin{equation}
	S=\begin{bmatrix}
		0&0&0&0&\cdots\\
		1&0&0&0&\cdots\\
		0&1&0&0&\cdots\\
		0&0&1&0&\cdots\\
		\vdots&\vdots&\vdots&\vdots&\ddots
	\end{bmatrix},\quad S^\dagger=\begin{bmatrix}
		0&1&0&0&\cdots\\
		0&0&1&0&\cdots\\
		0&0&0&1&\cdots\\
		0&0&0&0&\cdots\\
		\vdots&\vdots&\vdots&\vdots&\ddots
	\end{bmatrix}.\nonumber
\end{equation}
By taking the ansatz
\begin{equation}
	|\psi_{\k}\rangle=\sum_{i=0}^{\infty} \lambda^i|i\rangle\otimes|\xi\rangle,\quad |\lambda|<1
\end{equation}
for the boundary states, we solve the eigenvalue equation of Hamiltonian. In the bulk with $i\ge 1$, we have
\begin{widetext}
\begin{equation}
	\begin{bmatrix}
		0&t_1&0&0&0&t_2&0&0&t'_3e^{-ik_y}&0&0&0\\
		t_1&0&t_3&0&0&0&0&0&0&t'_2e^{-ik_y}\lambda&0&0\\
		0&t_3&0&t_2&0&t'_1\lambda&0&0&0&0&0&0\\
		0&0&t_2&0&t_1&0&0&0&0&0&0&t'_3\lambda\\
		0&0&0&t_1&0&t_3&t'_2&0&0&0&0&0\\
		t_2&0&t'_1\lambda^{-1}&0&t_3&0&0&0&0&0&0&0\\
		0&0&0&0&t'_2&0&0&t_3&0&0&0&t_1\\
		0&0&0&0&0&0&t_3&0&t_2&0&t'_1\lambda&0\\
		t'_3e^{ik_y}&0&0&0&0&0&0&t_2&0&t_1&0&0\\
		0&t'_2e^{ik_y}\lambda^{-1}&0&0&0&0&0&0&t_1&0&t_3&0\\
		0&0&0&0&0&0&0&t'_1\lambda^{-1}&0&t_3&0&t_2\\
		0&0&0&t'_3\lambda^{-1}&0&0&t_1&0&0&0&t_2&0
	\end{bmatrix}\begin{bmatrix}
		\xi_1\\
		\xi_2\\
		\xi_3\\
		\xi_4\\
		\xi_5\\
		\xi_6\\
		\xi_7\\
		\xi_8\\
		\xi_9\\
		\xi_{10}\\
		\xi_{11}\\
		\xi_{12}
	\end{bmatrix}=\mathcal{E}\begin{bmatrix}
		\xi_1\\
		\xi_2\\
		\xi_3\\
		\xi_4\\
		\xi_5\\
		\xi_6\\
		\xi_7\\
		\xi_8\\
		\xi_9\\
		\xi_{10}\\
		\xi_{11}\\
		\xi_{12}
	\end{bmatrix}\label{i>1}
\end{equation}
\end{widetext}
On the boundary with $i=0$, we have
\begin{widetext}
\begin{equation}
	\begin{bmatrix}
		0&t_1&0&0&0&t_2&0&0&t'_3e^{-ik_y}&0&0&0\\
		t_1&0&t_3&0&0&0&0&0&0&t'_2e^{-ik_y}\lambda&0&0\\
		0&t_3&0&t_2&0&t'_1\lambda&0&0&0&0&0&0\\
		0&0&t_2&0&t_1&0&0&0&0&0&0&t'_3\lambda\\
		0&0&0&t_1&0&t_3&t'_2&0&0&0&0&0\\
		t_2&0&0&0&t_3&0&0&0&0&0&0&0\\
		0&0&0&0&t'_2&0&0&t_3&0&0&0&t_1\\
		0&0&0&0&0&0&t_3&0&t_2&0&t'_1\lambda&0\\
		t'_3e^{ik_y}&0&0&0&0&0&0&t_2&0&t_1&0&0\\
		0&0&0&0&0&0&0&0&t_1&0&t_3&0\\
		0&0&0&0&0&0&0&0&0&t_3&0&t_2\\
		0&0&0&0&0&0&t_1&0&0&0&t_2&0
	\end{bmatrix}\begin{bmatrix}
		\xi_1\\
		\xi_2\\
		\xi_3\\
		\xi_4\\
		\xi_5\\
		\xi_6\\
		\xi_7\\
		\xi_8\\
		\xi_9\\
		\xi_{10}\\
		\xi_{11}\\
		\xi_{12}
	\end{bmatrix}=\mathcal{E}\begin{bmatrix}
		\xi_1\\
		\xi_2\\
		\xi_3\\
		\xi_4\\
		\xi_5\\
		\xi_6\\
		\xi_7\\
		\xi_8\\
		\xi_9\\
		\xi_{10}\\
		\xi_{11}\\
		\xi_{12}
	\end{bmatrix}\label{i=0}
\end{equation}
\end{widetext}
The difference of Eqs. \eqref{i>1} and \eqref{i=0} gives 
\begin{equation}
	\xi_2=\xi_3=\xi_4=\xi_8=0.
\end{equation}
Since the boundary effective Hamiltonian is given by
\begin{equation}
	\H_{\mathrm{eff}}=\langle\xi^i|\H|\xi^j\rangle,
\end{equation}
we obtain the equivalent form of boundary effective Hamiltonian, namely, by deleting the $2^{\mathrm{nd}},3^{\mathrm{rd}},4^{\mathrm{th}},8^{\mathrm{th}}$ rows and columns of the original Hamiltonian,
\begin{equation}
	\H_{k_y}=\begin{bmatrix}
		0&0&t_2&0&t'_3e^{-ik_y}&0&0&0\\
		0&0&t_3&t'_2&0&0&0&0\\
		t_2&t_3&0&0&0&0&0&0\\
		0&t'_2&0&0&0&0&0&t_1\\
		t'_3e^{ik_y}&0&0&0&0&t_1&0&0\\
		0&0&0&0&t_1&0&t_3&0\\
		0&0&0&0&0&t_3&0&t_2\\
		0&0&0&t_1&0&0&t_2&0
	\end{bmatrix}.
\end{equation}
Since the topological transitions must undergo a gapless state (or break symmetries, but we preserve symmetries here), we can derive the critical conditons of phase-transition point by letting the determinant of $\H_{k_y}$ being equal to $0$, namely,
\begin{equation}
	\begin{split}
		\mathrm{det}(\H_{k_y})&=t_1^2(t_2^4(t'_2)^2+t_3^2(t'_3)^2-2t_2^2t_3^2t'_2t_3\cos k_y)\\
		&\ge t_1^2(t_2^2t'_2-t_32t_3)^2\ge 0,\\
	\end{split}
\end{equation}
if and only if $k_y=0(k_y\in [-\pi,\pi])$, the first equal sign establishes in the first inequality. And if and only if $k_y=0$ and $a^2_2t'_2=t_3^2t'_3$, the second equal sign establishes in the second inequlity. Hence the condition of det$(\H_{k_y})=0$ having solutions is 
\begin{equation}
	a^2_2t'_2=t_3^2t'_3.\label{zigzag-condition}
\end{equation}
When the above equation holds, there exist gapless boundary states along the zigzag boundary, more explicitly, there is a Dirac point localized at $k_y=0$.

\subsection{The armchair boundary}

Following the same argument, we can obtain the boundary effective Hamiltonian along the periodic $x$ direction, namely, the armchair boundary,
\begin{widetext}
\begin{equation}
	\H_{k_x}=\left(
	\begin{array}{cccccccccc}
		0 & {t_1} & 0 & 0 & 0 & {t_2} & 0 & 0 & 0 & 0 \\
		{t_1} & 0 & {t_3} & 0 & 0 & 0 & 0 & 0 & 0 & 0 \\
		0 & {t_3} & 0 & {t_2} & 0 & {t'_1} e^{i {k_x}} & 0 & 0 & 0 & 0 \\
		0 & 0 & {t_2} & 0 & {t_1} & 0 & 0 & 0 & 0 & {t'_3} e^{i {k_x}} \\
		0 & 0 & 0 & {t_1} & 0 & {t_3} & {t'_2} & 0 & 0 & 0 \\
		{t_2} & 0 & {t'_1} e^{-i {k_x}} & 0 & {t_3} & 0 & 0 & 0 & 0 & 0 \\
		0 & 0 & 0 & 0 & {t'_2} & 0 & 0 & {t_3} & 0 & {t_1} \\
		0 & 0 & 0 & 0 & 0 & 0 & {t_3} & 0 & {t'_1} e^{i {k_x}} & 0 \\
		0 & 0 & 0 & 0 & 0 & 0 & 0 & {t'_1} e^{-i {k_x}} & 0 & {t_2} \\
		0 & 0 & 0 & {t'_3} e^{-i {k_x}} & 0 & 0 & {t_1} & 0 & {t_2} & 0 \\
	\end{array}
	\right).
\end{equation}
\end{widetext}
Thus we have
\begin{widetext}
\begin{equation}
	\begin{split}
		\mathrm{det}(\H_{k_x})=&(t_2^2t_3^3+t_1^2(t'_1)^2-2t_1t_2t_3t'_1\cos k_x)\\
		&\cdot ({t_1}^4 {t'_1}^2+4 {t_1}^2 {t_2}^2 {t_3}^2-2 {t_1}^2 {t'_1} \cos {k_x} (2 {t_1} {t_2} {t_3}+{t'_1} {t'_2} {t'_3})+4 {t_1} {t_2} {t_3} {t'_1} {t'_2} {t'_3}+{t'_1}^2 {t'_2}^2 {t'_3}^2)\\
		\ge &(t_2t_3-t_1t'_1)^
		2(t_1^2t'_1-(2t_1t_2t_3+t'_1t'_2t'_3))^2\ge 0.
	\end{split}
\end{equation}
\end{widetext}
if and only if $k_x=0(k_y\in [-\pi,\pi])$, the first equal sign establishes in the first inequality. And if and only if $k_x=0$ and $t_2t_3=t_1t'_1$ or $t_1^2t'_1=2t_1t_2t_3+t'_1t'_2t'_3$, the second equal sign establishes in the second inequlity. However, the condition $t_1^2t'_1=2t_1t_2t_3+t'_1t'_2t'_3$ corresponds the trivial topological phases.
Hence the condition of det$(\H_{k_y})=0$ having solutions is 
\begin{equation}
	t_2t_3=t_1t'_1.\label{armchair-condition}
\end{equation}
It is noted that the relation is same as that of rohmbic sample. 
When the above equation holds, there exist gapless boundary states along the armchair boundary. More explicitly, there is a Dirac point localized at $k_x=0$.
\begin{figure}
	\includegraphics[scale=0.14]{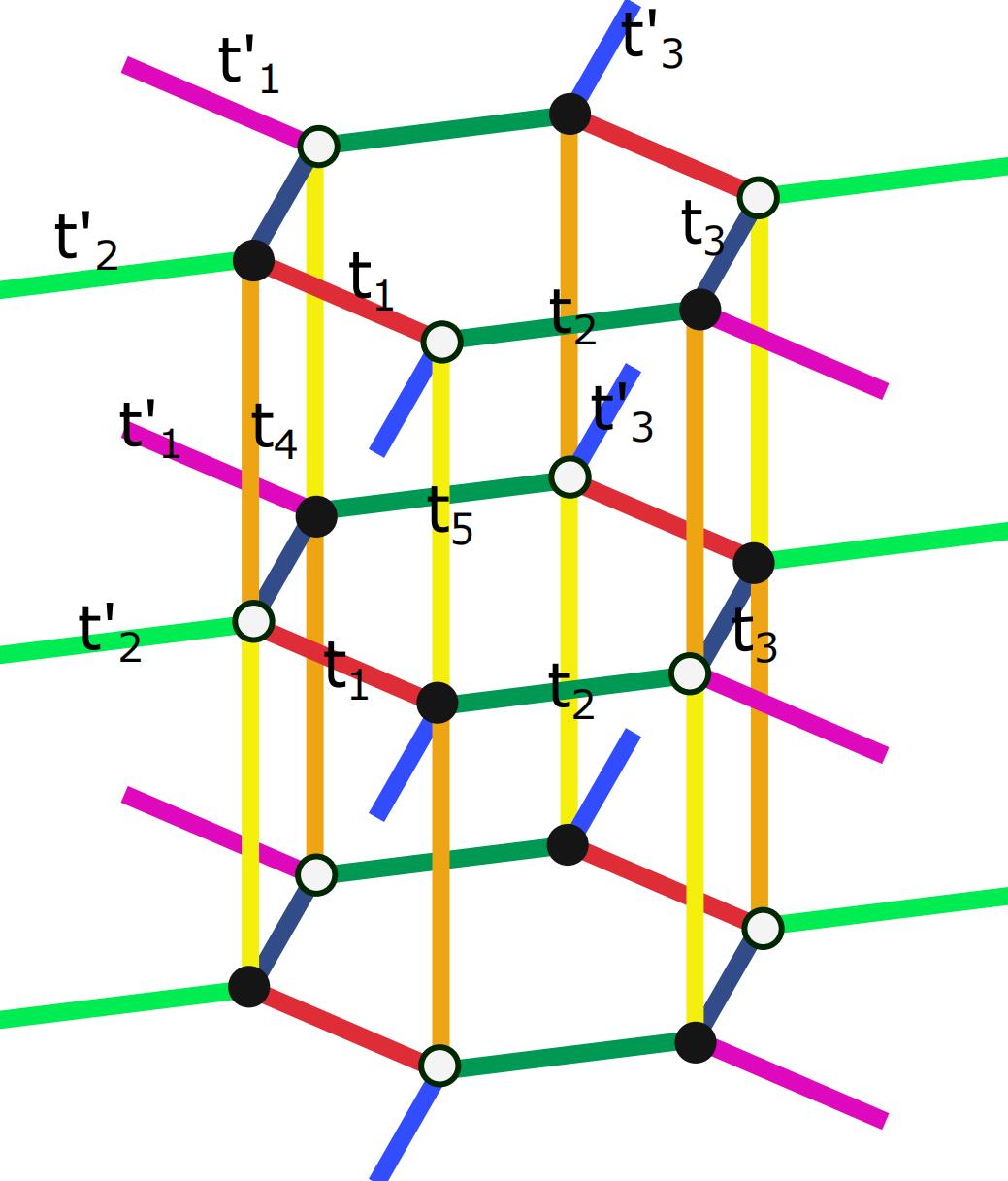}
	\caption{The lattice structure of 3D Tagaki insulator, which is formed by satcking the 2D honeycomb lattice with alternative interlayer hoppings. The hopping between layers is colored by  orange and yellow.}	\label{3Dlattice}
\end{figure}

\section{3D honeycomb model}\label{3dmodel}
We construct a 3D model by stacking 2D honeycomb Takagi insulator in a manner of stageer as shown in Fig.\ref{3Dlattice}. Here $t_{i}=t,t'_{i}=T$, the Hamiltonian is given by 
\begin{equation}
	H(k)=\left[\begin{array}{cc}
		0 & Q(k) \\
		Q^{\dagger}(k) & 0
	\end{array}\right], Q=\left[\begin{array}{cc}
		A(k) & e^{i k_{z}} B(k) \\
		B(k) & A^{*}(k)
	\end{array}\right],
\end{equation}
where $B(k)=\left(t_{4}+t_{5} e^{-i k_{z}}\right) I_{3}$ 
and 
\begin{equation}
	A(k)=\left[\begin{array}{ccc}
		T e^{i k \cdot \mathbf{a}_{1}} & t & t \\
		t & T e^{i k \cdot \mathbf{a}_{2}} & t \\
		t & t & T e^{i k \cdot \mathbf{a}_{3}}
	\end{array}\right].
\end{equation}
To get corner zero-modes of hexagonal prisms sample in main text, we set parameters as  $t=1,T=3,t_4=0.5,t_6=0.8.$

\section{Different boundary modes}\label{different_boundary}

\begin{figure*}[t]
	\includegraphics[scale=0.33]{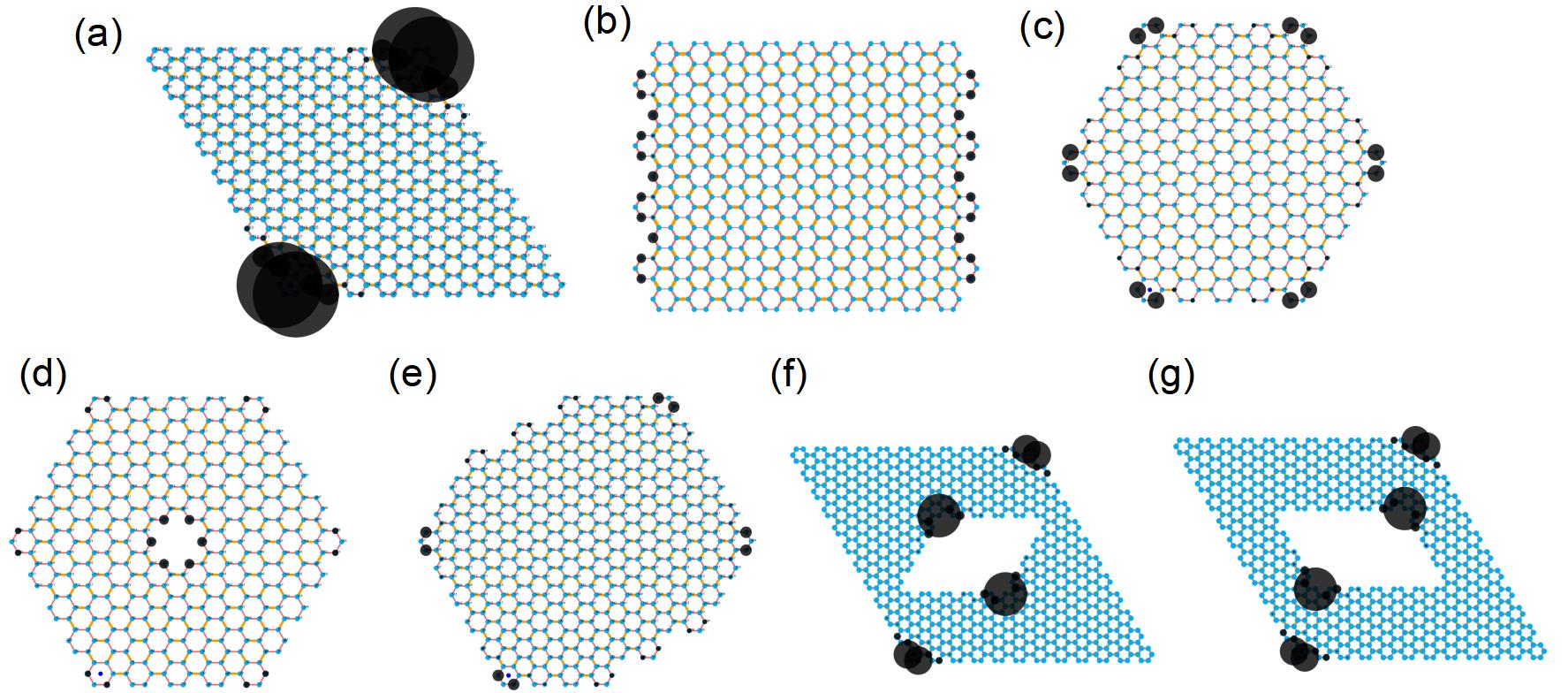}
	\caption{Different boundary modes for arbitary-shaped sample keeping $PT$ symmetry. With Same parameter but different open boundary conditions can bring different corner states and even helical edge states. The parameters for all samples are set as  $t_{1}=t_{2}=t_{3}<t'_{1}=t'_{2}=t'_{3}$. }	\label{boundary_modes}
\end{figure*}

As proposed in maintext, by cutting different-shaped samples that preserve the $\P\T$ symmetry, without changing the parameters, the system can possess distinguishable boundary-modes.
\begin{figure}
	\includegraphics[scale=0.25]{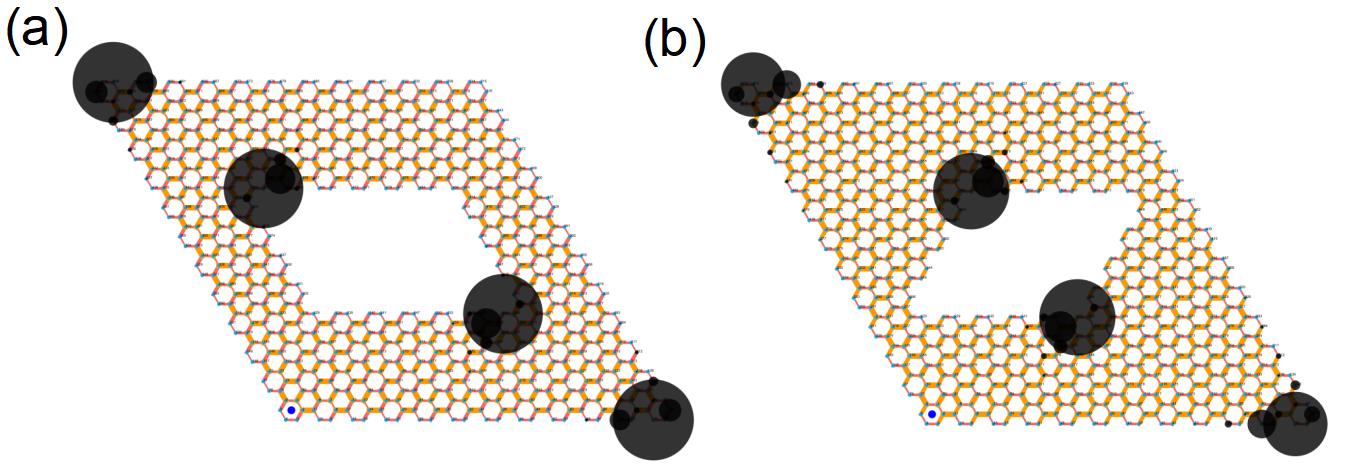}
	\caption{Rhombic sample with a hollow rhombus. The zero-modes can locate on different corner by cutting different shape with  invariant parameters as (a) and (b).}	
	\label{rhombicAngle}
\end{figure}
Here we take some representatived example to show that result, which can be seen in Fig.\ref{boundary_modes} and \ref{rhombicAngle}. It is noted that, as shown in Fig.\ref{rhombicAngle}, we can tune the position of corner zero-modes by cutting different gometric shape. Specificly, without changing parameters the local zero-modes can locate on different $\P\T$ symmetry-related corners (acute angle or obtuse angle).

\bibliographystyle{apsrev4-1}
\bibliography{ref-SSH2D}

\begin{thebibliography}{51}%
\makeatletter
\providecommand \@ifxundefined [1]{%
 \@ifx{#1\undefined}
}%
\providecommand \@ifnum [1]{%
 \ifnum #1\expandafter \@firstoftwo
 \else \expandafter \@secondoftwo
 \fi
}%
\providecommand \@ifx [1]{%
 \ifx #1\expandafter \@firstoftwo
 \else \expandafter \@secondoftwo
 \fi
}%
\providecommand \natexlab [1]{#1}%
\providecommand \enquote  [1]{``#1''}%
\providecommand \bibnamefont  [1]{#1}%
\providecommand \bibfnamefont [1]{#1}%
\providecommand \citenamefont [1]{#1}%
\providecommand \href@noop [0]{\@secondoftwo}%
\providecommand \href [0]{\begingroup \@sanitize@url \@href}%
\providecommand \@href[1]{\@@startlink{#1}\@@href}%
\providecommand \@@href[1]{\endgroup#1\@@endlink}%
\providecommand \@sanitize@url [0]{\catcode `\\12\catcode `\$12\catcode
  `\&12\catcode `\#12\catcode `\^12\catcode `\_12\catcode `\%12\relax}%
\providecommand \@@startlink[1]{}%
\providecommand \@@endlink[0]{}%
\providecommand \url  [0]{\begingroup\@sanitize@url \@url }%
\providecommand \@url [1]{\endgroup\@href {#1}{\urlprefix }}%
\providecommand \urlprefix  [0]{URL }%
\providecommand \Eprint [0]{\href }%
\providecommand \doibase [0]{http://dx.doi.org/}%
\providecommand \selectlanguage [0]{\@gobble}%
\providecommand \bibinfo  [0]{\@secondoftwo}%
\providecommand \bibfield  [0]{\@secondoftwo}%
\providecommand \translation [1]{[#1]}%
\providecommand \BibitemOpen [0]{}%
\providecommand \bibitemStop [0]{}%
\providecommand \bibitemNoStop [0]{.\EOS\space}%
\providecommand \EOS [0]{\spacefactor3000\relax}%
\providecommand \BibitemShut  [1]{\csname bibitem#1\endcsname}%
\let\auto@bib@innerbib\@empty
\bibitem [{\citenamefont {Volovik}(2003)}]{Volovik:book}%
  \BibitemOpen
  \bibfield  {author} {\bibinfo {author} {\bibfnamefont {G.~E.}\ \bibnamefont
  {Volovik}},\ }\href@noop {} {\emph {\bibinfo {title} {Universe in a helium
  droplet}}}\ (\bibinfo  {publisher} {Oxford University Press, Oxford UK},\
  \bibinfo {year} {2003})\BibitemShut {NoStop}%
\bibitem [{\citenamefont {Hasan}\ and\ \citenamefont {Kane}(2010)}]{Kane-RMP}%
  \BibitemOpen
  \bibfield  {author} {\bibinfo {author} {\bibfnamefont {M.~Z.}\ \bibnamefont
  {Hasan}}\ and\ \bibinfo {author} {\bibfnamefont {C.~L.}\ \bibnamefont
  {Kane}},\ }\href {\doibase 10.1103/RevModPhys.82.3045} {\bibfield  {journal}
  {\bibinfo  {journal} {Rev. Mod. Phys.}\ }\textbf {\bibinfo {volume} {82}},\
  \bibinfo {pages} {3045} (\bibinfo {year} {2010})}\BibitemShut {NoStop}%
\bibitem [{\citenamefont {Qi}\ and\ \citenamefont {Zhang}(2011)}]{XLQi-RMP}%
  \BibitemOpen
  \bibfield  {author} {\bibinfo {author} {\bibfnamefont {X.-L.}\ \bibnamefont
  {Qi}}\ and\ \bibinfo {author} {\bibfnamefont {S.-C.}\ \bibnamefont {Zhang}},\
  }\href {\doibase 10.1103/RevModPhys.83.1057} {\bibfield  {journal} {\bibinfo
  {journal} {Rev. Mod. Phys.}\ }\textbf {\bibinfo {volume} {83}},\ \bibinfo
  {pages} {1057} (\bibinfo {year} {2011})}\BibitemShut {NoStop}%
\bibitem [{\citenamefont {Fu}(2011)}]{FuLiang2011prl}%
  \BibitemOpen
  \bibfield  {author} {\bibinfo {author} {\bibfnamefont {L.}~\bibnamefont
  {Fu}},\ }\href {\doibase 10.1103/PhysRevLett.106.106802} {\bibfield
  {journal} {\bibinfo  {journal} {Phys. Rev. Lett.}\ }\textbf {\bibinfo
  {volume} {106}},\ \bibinfo {pages} {106802} (\bibinfo {year}
  {2011})}\BibitemShut {NoStop}%
\bibitem [{\citenamefont {Chiu}\ \emph {et~al.}(2016)\citenamefont {Chiu},
  \citenamefont {Teo}, \citenamefont {Ascnyder},\ and\ \citenamefont
  {Ryu}}]{ShinseiRyu-RMP}%
  \BibitemOpen
  \bibfield  {author} {\bibinfo {author} {\bibfnamefont {C.-K.}\ \bibnamefont
  {Chiu}}, \bibinfo {author} {\bibfnamefont {J.~C.~Y.}\ \bibnamefont {Teo}},
  \bibinfo {author} {\bibfnamefont {A.~P.}\ \bibnamefont {Ascnyder}}, \ and\
  \bibinfo {author} {\bibfnamefont {S.}~\bibnamefont {Ryu}},\ }\href {\doibase
  10.1103/RevModPhys.88.035005} {\bibfield  {journal} {\bibinfo  {journal}
  {Rev. Mod. Phys.}\ }\textbf {\bibinfo {volume} {88}},\ \bibinfo {pages}
  {035005} (\bibinfo {year} {2016})}\BibitemShut {NoStop}%
\bibitem [{\citenamefont {Kruthoff}\ \emph {et~al.}(2017)\citenamefont
  {Kruthoff}, \citenamefont {de~Boer}, \citenamefont {van Wezel}, \citenamefont
  {Kane},\ and\ \citenamefont {Slager}}]{Kruthoff2017prx}%
  \BibitemOpen
  \bibfield  {author} {\bibinfo {author} {\bibfnamefont {J.}~\bibnamefont
  {Kruthoff}}, \bibinfo {author} {\bibfnamefont {J.}~\bibnamefont {de~Boer}},
  \bibinfo {author} {\bibfnamefont {J.}~\bibnamefont {van Wezel}}, \bibinfo
  {author} {\bibfnamefont {C.~L.}\ \bibnamefont {Kane}}, \ and\ \bibinfo
  {author} {\bibfnamefont {R.-J.}\ \bibnamefont {Slager}},\ }\href {\doibase
  10.1103/PhysRevX.7.041069} {\bibfield  {journal} {\bibinfo  {journal} {Phys.
  Rev. X}\ }\textbf {\bibinfo {volume} {7}},\ \bibinfo {pages} {041069}
  (\bibinfo {year} {2017})}\BibitemShut {NoStop}%
\bibitem [{\citenamefont {Benalcazar}\ \emph {et~al.}(2017)\citenamefont
  {Benalcazar}, \citenamefont {Bernevig},\ and\ \citenamefont
  {Hughes}}]{Benalcazar2017prb}%
  \BibitemOpen
  \bibfield  {author} {\bibinfo {author} {\bibfnamefont {W.~A.}\ \bibnamefont
  {Benalcazar}}, \bibinfo {author} {\bibfnamefont {B.~A.}\ \bibnamefont
  {Bernevig}}, \ and\ \bibinfo {author} {\bibfnamefont {T.~L.}\ \bibnamefont
  {Hughes}},\ }\href {\doibase 10.1103/PhysRevB.96.245115} {\bibfield
  {journal} {\bibinfo  {journal} {Phys. Rev. B}\ }\textbf {\bibinfo {volume}
  {96}},\ \bibinfo {pages} {245115} (\bibinfo {year} {2017})}\BibitemShut
  {NoStop}%
\bibitem [{\citenamefont {Liu}\ \emph {et~al.}(2019)\citenamefont {Liu},
  \citenamefont {Deng},\ and\ \citenamefont {Wakabayashi}}]{LiuFeng2019prl}%
  \BibitemOpen
  \bibfield  {author} {\bibinfo {author} {\bibfnamefont {F.}~\bibnamefont
  {Liu}}, \bibinfo {author} {\bibfnamefont {H.-Y.}\ \bibnamefont {Deng}}, \
  and\ \bibinfo {author} {\bibfnamefont {K.}~\bibnamefont {Wakabayashi}},\
  }\href {\doibase 10.1103/PhysRevLett.122.086804} {\bibfield  {journal}
  {\bibinfo  {journal} {Phys. Rev. Lett.}\ }\textbf {\bibinfo {volume} {122}},\
  \bibinfo {pages} {086804} (\bibinfo {year} {2019})}\BibitemShut {NoStop}%
\bibitem [{\citenamefont {Xie}\ \emph {et~al.}(2021)\citenamefont {Xie},
  \citenamefont {Wang}, \citenamefont {Zhang}, \citenamefont {Zhan},
  \citenamefont {Jiang}, \citenamefont {Lu},\ and\ \citenamefont
  {Chen}}]{xie2021higher}%
  \BibitemOpen
  \bibfield  {author} {\bibinfo {author} {\bibfnamefont {B.}~\bibnamefont
  {Xie}}, \bibinfo {author} {\bibfnamefont {H.-X.}\ \bibnamefont {Wang}},
  \bibinfo {author} {\bibfnamefont {X.}~\bibnamefont {Zhang}}, \bibinfo
  {author} {\bibfnamefont {P.}~\bibnamefont {Zhan}}, \bibinfo {author}
  {\bibfnamefont {J.-H.}\ \bibnamefont {Jiang}}, \bibinfo {author}
  {\bibfnamefont {M.}~\bibnamefont {Lu}}, \ and\ \bibinfo {author}
  {\bibfnamefont {Y.}~\bibnamefont {Chen}},\ }\href@noop {} {\bibfield
  {journal} {\bibinfo  {journal} {Nat. Rev. Phys.}\ }\textbf {\bibinfo {volume}
  {3}},\ \bibinfo {pages} {520} (\bibinfo {year} {2021})}\BibitemShut {NoStop}%
\bibitem [{\citenamefont {Atiyah}(1966)}]{Atiyah-KR}%
  \BibitemOpen
  \bibfield  {author} {\bibinfo {author} {\bibfnamefont {M.~F.}\ \bibnamefont
  {Atiyah}},\ }\href {\doibase 10.1093/qmath/17.1.367} {\bibfield  {journal}
  {\bibinfo  {journal} {The Quarterly Journal of Mathematics}\ }\textbf
  {\bibinfo {volume} {17}},\ \bibinfo {pages} {367} (\bibinfo {year}
  {1966})}\BibitemShut {NoStop}%
\bibitem [{\citenamefont {Kitaev}(2010)}]{Kitaev2009AIP}%
  \BibitemOpen
  \bibfield  {author} {\bibinfo {author} {\bibfnamefont {A.}~\bibnamefont
  {Kitaev}},\ }\href {\doibase 10.1063/1.3149495} {\bibfield  {journal}
  {\bibinfo  {journal} {AIP Conference Proceedings}\ }\textbf {\bibinfo
  {volume} {1134}},\ \bibinfo {pages} {22} (\bibinfo {year}
  {2010})}\BibitemShut {NoStop}%
\bibitem [{\citenamefont {{Schnyder}}\ \emph {et~al.}(2008)\citenamefont
  {{Schnyder}}, \citenamefont {{Ryu}}, \citenamefont {{Furusaki}},\ and\
  \citenamefont {{Ludwig}}}]{Schnyder2008}%
  \BibitemOpen
  \bibfield  {author} {\bibinfo {author} {\bibfnamefont {A.~P.}\ \bibnamefont
  {{Schnyder}}}, \bibinfo {author} {\bibfnamefont {S.}~\bibnamefont {{Ryu}}},
  \bibinfo {author} {\bibfnamefont {A.}~\bibnamefont {{Furusaki}}}, \ and\
  \bibinfo {author} {\bibfnamefont {A.~W.~W.}\ \bibnamefont {{Ludwig}}},\
  }\href {\doibase 10.1103/PhysRevB.78.195125} {\bibfield  {journal} {\bibinfo
  {journal} {\prb}\ }\textbf {\bibinfo {volume} {78}},\ \bibinfo {eid} {195125}
  (\bibinfo {year} {2008})}\BibitemShut {NoStop}%
\bibitem [{\citenamefont {Altland}\ and\ \citenamefont
  {Zirnbauer}(1997)}]{AZ-Classification}%
  \BibitemOpen
  \bibfield  {author} {\bibinfo {author} {\bibfnamefont {A.}~\bibnamefont
  {Altland}}\ and\ \bibinfo {author} {\bibfnamefont {M.~R.}\ \bibnamefont
  {Zirnbauer}},\ }\href {\doibase 10.1103/PhysRevB.55,1142} {\bibfield
  {journal} {\bibinfo  {journal} {Phys. Rev. B}\ }\textbf {\bibinfo {volume}
  {55}},\ \bibinfo {pages} {1142} (\bibinfo {year} {1997})}\BibitemShut
  {NoStop}%
\bibitem [{\citenamefont {Ho\ifmmode~\check{r}\else
  \v{r}\fi{}ava}(2005)}]{HoravaPRL05}%
  \BibitemOpen
  \bibfield  {author} {\bibinfo {author} {\bibfnamefont {P.}~\bibnamefont
  {Ho\ifmmode~\check{r}\else \v{r}\fi{}ava}},\ }\href {\doibase
  10.1103/PhysRevLett.95.016405} {\bibfield  {journal} {\bibinfo  {journal}
  {Phys. Rev. Lett.}\ }\textbf {\bibinfo {volume} {95}},\ \bibinfo {pages}
  {016405} (\bibinfo {year} {2005})}\BibitemShut {NoStop}%
\bibitem [{\citenamefont {Zhao}\ and\ \citenamefont
  {Wang}(2014)}]{ZhaoYXWang14Septprb}%
  \BibitemOpen
  \bibfield  {author} {\bibinfo {author} {\bibfnamefont {Y.~X.}\ \bibnamefont
  {Zhao}}\ and\ \bibinfo {author} {\bibfnamefont {Z.~D.}\ \bibnamefont
  {Wang}},\ }\href {\doibase 10.1103/PhysRevB.89.075111} {\bibfield  {journal}
  {\bibinfo  {journal} {Phys. Rev. B}\ }\textbf {\bibinfo {volume} {89}},\
  \bibinfo {pages} {075111} (\bibinfo {year} {2014})}\BibitemShut {NoStop}%
\bibitem [{\citenamefont {Ryu}\ \emph {et~al.}(2010)\citenamefont {Ryu},
  \citenamefont {Schnyder}, \citenamefont {Furusaki},\ and\ \citenamefont
  {Ludwig}}]{ShinseiRyu2010NJP}%
  \BibitemOpen
  \bibfield  {author} {\bibinfo {author} {\bibfnamefont {S.}~\bibnamefont
  {Ryu}}, \bibinfo {author} {\bibfnamefont {A.~P.}\ \bibnamefont {Schnyder}},
  \bibinfo {author} {\bibfnamefont {A.}~\bibnamefont {Furusaki}}, \ and\
  \bibinfo {author} {\bibfnamefont {A.~W.~W.}\ \bibnamefont {Ludwig}},\ }\href
  {\doibase 10.1088/1367-2630/12/6/065010} {\bibfield  {journal} {\bibinfo
  {journal} {New Journal of Physics}\ }\textbf {\bibinfo {volume} {12}},\
  \bibinfo {pages} {065010} (\bibinfo {year} {2010})}\BibitemShut {NoStop}%
\bibitem [{\citenamefont {Matsuura}\ \emph {et~al.}(2013)\citenamefont
  {Matsuura}, \citenamefont {Chang}, \citenamefont {Schnyder},\ and\
  \citenamefont {Ryu}}]{matsuura2013protected}%
  \BibitemOpen
  \bibfield  {author} {\bibinfo {author} {\bibfnamefont {S.}~\bibnamefont
  {Matsuura}}, \bibinfo {author} {\bibfnamefont {P.-Y.}\ \bibnamefont {Chang}},
  \bibinfo {author} {\bibfnamefont {A.~P.}\ \bibnamefont {Schnyder}}, \ and\
  \bibinfo {author} {\bibfnamefont {S.}~\bibnamefont {Ryu}},\ }\href@noop {}
  {\bibfield  {journal} {\bibinfo  {journal} {New Journal of Physics}\ }\textbf
  {\bibinfo {volume} {15}},\ \bibinfo {pages} {065001} (\bibinfo {year}
  {2013})}\BibitemShut {NoStop}%
\bibitem [{\citenamefont {Zhao}\ and\ \citenamefont
  {Wang}(2013)}]{ZhaoYXWang13prl}%
  \BibitemOpen
  \bibfield  {author} {\bibinfo {author} {\bibfnamefont {Y.~X.}\ \bibnamefont
  {Zhao}}\ and\ \bibinfo {author} {\bibfnamefont {Z.~D.}\ \bibnamefont
  {Wang}},\ }\href {\doibase 10.1103/PhysRevLett.110.240404} {\bibfield
  {journal} {\bibinfo  {journal} {Phys. Rev. Lett.}\ }\textbf {\bibinfo
  {volume} {110}},\ \bibinfo {pages} {240404} (\bibinfo {year}
  {2013})}\BibitemShut {NoStop}%
\bibitem [{\citenamefont {Chiu}\ and\ \citenamefont
  {Schnyder}(2014)}]{Chiu2014}%
  \BibitemOpen
  \bibfield  {author} {\bibinfo {author} {\bibfnamefont {C.-K.}\ \bibnamefont
  {Chiu}}\ and\ \bibinfo {author} {\bibfnamefont {A.~P.}\ \bibnamefont
  {Schnyder}},\ }\href {\doibase 10.1103/PhysRevB.90.205136} {\bibfield
  {journal} {\bibinfo  {journal} {Phys. Rev. B}\ }\textbf {\bibinfo {volume}
  {90}},\ \bibinfo {pages} {205136} (\bibinfo {year} {2014})}\BibitemShut
  {NoStop}%
\bibitem [{\citenamefont {Shiozaki}\ and\ \citenamefont
  {Sato}(2014)}]{Sato2014PRB}%
  \BibitemOpen
  \bibfield  {author} {\bibinfo {author} {\bibfnamefont {K.}~\bibnamefont
  {Shiozaki}}\ and\ \bibinfo {author} {\bibfnamefont {M.}~\bibnamefont
  {Sato}},\ }\href {\doibase 10.1103/PhysRevB.90.165114} {\bibfield  {journal}
  {\bibinfo  {journal} {Phys. Rev. B}\ }\textbf {\bibinfo {volume} {90}},\
  \bibinfo {pages} {165114} (\bibinfo {year} {2014})}\BibitemShut {NoStop}%
\bibitem [{\citenamefont {Zhao}\ \emph {et~al.}(2016)\citenamefont {Zhao},
  \citenamefont {Schnyder},\ and\ \citenamefont {Wang}}]{ZhaoWang16Aprprl}%
  \BibitemOpen
  \bibfield  {author} {\bibinfo {author} {\bibfnamefont {X.}~\bibnamefont
  {Zhao}}, \bibinfo {author} {\bibfnamefont {A.~P.}\ \bibnamefont {Schnyder}},
  \ and\ \bibinfo {author} {\bibfnamefont {Z.~D.}\ \bibnamefont {Wang}},\
  }\href {\doibase 10.1103/PhysRevLett.116.156402} {\bibfield  {journal}
  {\bibinfo  {journal} {Phys. Rev. Lett.}\ }\textbf {\bibinfo {volume} {116}},\
  \bibinfo {pages} {156402} (\bibinfo {year} {2016})}\BibitemShut {NoStop}%
\bibitem [{\citenamefont {Zhao}\ and\ \citenamefont
  {Lu}(2017)}]{ZhaoLu17Aprprl}%
  \BibitemOpen
  \bibfield  {author} {\bibinfo {author} {\bibfnamefont {Y.~X.}\ \bibnamefont
  {Zhao}}\ and\ \bibinfo {author} {\bibfnamefont {Y.}~\bibnamefont {Lu}},\
  }\href {\doibase 10.1103/PhysRevLett.118.056401} {\bibfield  {journal}
  {\bibinfo  {journal} {Phys. Rev. Lett.}\ }\textbf {\bibinfo {volume} {118}},\
  \bibinfo {pages} {056401} (\bibinfo {year} {2017})}\BibitemShut {NoStop}%
\bibitem [{\citenamefont {Ahn}\ \emph {et~al.}(2019)\citenamefont {Ahn},
  \citenamefont {Park},\ and\ \citenamefont {Yang}}]{B-J-Yang19APRPRX}%
  \BibitemOpen
  \bibfield  {author} {\bibinfo {author} {\bibfnamefont {J.}~\bibnamefont
  {Ahn}}, \bibinfo {author} {\bibfnamefont {S.}~\bibnamefont {Park}}, \ and\
  \bibinfo {author} {\bibfnamefont {B.-J.}\ \bibnamefont {Yang}},\ }\href
  {\doibase 10.1103/PhysRevX.9.021013} {\bibfield  {journal} {\bibinfo
  {journal} {Phys. Rev. X}\ } (\bibinfo {year} {2019}),\
  10.1103/PhysRevX.9.021013}\BibitemShut {NoStop}%
\bibitem [{\citenamefont {Timm}\ \emph {et~al.}(2017)\citenamefont {Timm},
  \citenamefont {Schnyder}, \citenamefont {Agterberg},\ and\ \citenamefont
  {Brydon}}]{timm2017inflated}%
  \BibitemOpen
  \bibfield  {author} {\bibinfo {author} {\bibfnamefont {C.}~\bibnamefont
  {Timm}}, \bibinfo {author} {\bibfnamefont {A.~P.}\ \bibnamefont {Schnyder}},
  \bibinfo {author} {\bibfnamefont {D.~F.}\ \bibnamefont {Agterberg}}, \ and\
  \bibinfo {author} {\bibfnamefont {P.~M.~R.}\ \bibnamefont {Brydon}},\ }\href
  {\doibase 10.1103/PhysRevB.96.094526} {\bibfield  {journal} {\bibinfo
  {journal} {Phys. Rev. B}\ }\textbf {\bibinfo {volume} {96}},\ \bibinfo
  {pages} {094526} (\bibinfo {year} {2017})}\BibitemShut {NoStop}%
\bibitem [{\citenamefont {Yu}\ \emph {et~al.}(2021)\citenamefont {Yu},
  \citenamefont {Kennes}, \citenamefont {Rubio},\ and\ \citenamefont
  {Sentef}}]{PhysRevLett.127.127001}%
  \BibitemOpen
  \bibfield  {author} {\bibinfo {author} {\bibfnamefont {T.}~\bibnamefont
  {Yu}}, \bibinfo {author} {\bibfnamefont {D.~M.}\ \bibnamefont {Kennes}},
  \bibinfo {author} {\bibfnamefont {A.}~\bibnamefont {Rubio}}, \ and\ \bibinfo
  {author} {\bibfnamefont {M.~A.}\ \bibnamefont {Sentef}},\ }\href {\doibase
  10.1103/PhysRevLett.127.127001} {\bibfield  {journal} {\bibinfo  {journal}
  {Phys. Rev. Lett.}\ }\textbf {\bibinfo {volume} {127}},\ \bibinfo {pages}
  {127001} (\bibinfo {year} {2021})}\BibitemShut {NoStop}%
\bibitem [{\citenamefont {Tomonaga}\ \emph {et~al.}(2021)\citenamefont
  {Tomonaga}, \citenamefont {Mukai}, \citenamefont {Yoshihara},\ and\
  \citenamefont {Tsai}}]{tomonaga2021quasiparticle}%
  \BibitemOpen
  \bibfield  {author} {\bibinfo {author} {\bibfnamefont {A.}~\bibnamefont
  {Tomonaga}}, \bibinfo {author} {\bibfnamefont {H.}~\bibnamefont {Mukai}},
  \bibinfo {author} {\bibfnamefont {F.}~\bibnamefont {Yoshihara}}, \ and\
  \bibinfo {author} {\bibfnamefont {J.~S.}\ \bibnamefont {Tsai}},\ }\href
  {\doibase 10.1103/PhysRevB.104.224509} {\bibfield  {journal} {\bibinfo
  {journal} {Phys. Rev. B}\ }\textbf {\bibinfo {volume} {104}},\ \bibinfo
  {pages} {224509} (\bibinfo {year} {2021})}\BibitemShut {NoStop}%
\bibitem [{\citenamefont {Lapp}\ \emph {et~al.}(2020)\citenamefont {Lapp},
  \citenamefont {B\"orner},\ and\ \citenamefont {Timm}}]{lapp2020experimental}%
  \BibitemOpen
  \bibfield  {author} {\bibinfo {author} {\bibfnamefont {C.~J.}\ \bibnamefont
  {Lapp}}, \bibinfo {author} {\bibfnamefont {G.}~\bibnamefont {B\"orner}}, \
  and\ \bibinfo {author} {\bibfnamefont {C.}~\bibnamefont {Timm}},\ }\href
  {\doibase 10.1103/PhysRevB.101.024505} {\bibfield  {journal} {\bibinfo
  {journal} {Phys. Rev. B}\ }\textbf {\bibinfo {volume} {101}},\ \bibinfo
  {pages} {024505} (\bibinfo {year} {2020})}\BibitemShut {NoStop}%
\bibitem [{\citenamefont {Zhang}\ \emph {et~al.}(2013)\citenamefont {Zhang},
  \citenamefont {Kane},\ and\ \citenamefont {Mele}}]{Mele2013prl}%
  \BibitemOpen
  \bibfield  {author} {\bibinfo {author} {\bibfnamefont {F.}~\bibnamefont
  {Zhang}}, \bibinfo {author} {\bibfnamefont {C.~L.}\ \bibnamefont {Kane}}, \
  and\ \bibinfo {author} {\bibfnamefont {E.~J.}\ \bibnamefont {Mele}},\ }\href
  {\doibase 10.1103/PhysRevLett.110.046404} {\bibfield  {journal} {\bibinfo
  {journal} {Phys. Rev. Lett.}\ }\textbf {\bibinfo {volume} {110}},\ \bibinfo
  {pages} {046404} (\bibinfo {year} {2013})}\BibitemShut {NoStop}%
\bibitem [{\citenamefont {Yang}\ \emph {et~al.}(2015)\citenamefont {Yang},
  \citenamefont {Gao}, \citenamefont {Shi}, \citenamefont {Lin}, \citenamefont
  {Gao}, \citenamefont {Chong},\ and\ \citenamefont {Zhang}}]{YangZJ2015prl}%
  \BibitemOpen
  \bibfield  {author} {\bibinfo {author} {\bibfnamefont {Z.}~\bibnamefont
  {Yang}}, \bibinfo {author} {\bibfnamefont {F.}~\bibnamefont {Gao}}, \bibinfo
  {author} {\bibfnamefont {X.}~\bibnamefont {Shi}}, \bibinfo {author}
  {\bibfnamefont {X.}~\bibnamefont {Lin}}, \bibinfo {author} {\bibfnamefont
  {Z.}~\bibnamefont {Gao}}, \bibinfo {author} {\bibfnamefont {Y.}~\bibnamefont
  {Chong}}, \ and\ \bibinfo {author} {\bibfnamefont {B.}~\bibnamefont
  {Zhang}},\ }\href {\doibase 10.1103/PhysRevLett.114.114301} {\bibfield
  {journal} {\bibinfo  {journal} {Phys. Rev. Lett.}\ }\textbf {\bibinfo
  {volume} {114}},\ \bibinfo {pages} {114301} (\bibinfo {year}
  {2015})}\BibitemShut {NoStop}%
\bibitem [{\citenamefont {Imhof}\ \emph {et~al.}(2018)\citenamefont {Imhof},
  \citenamefont {Berger}, \citenamefont {Bayer}, \citenamefont {Brehm},
  \citenamefont {Molenkamp}, \citenamefont {Kiessling}, \citenamefont
  {Schindler}, \citenamefont {Lee}, \citenamefont {Greiter}, \citenamefont
  {Neupert},\ and\ \citenamefont {Thomale}}]{Ronny_2018np}%
  \BibitemOpen
  \bibfield  {author} {\bibinfo {author} {\bibfnamefont {S.}~\bibnamefont
  {Imhof}}, \bibinfo {author} {\bibfnamefont {C.}~\bibnamefont {Berger}},
  \bibinfo {author} {\bibfnamefont {F.}~\bibnamefont {Bayer}}, \bibinfo
  {author} {\bibfnamefont {J.}~\bibnamefont {Brehm}}, \bibinfo {author}
  {\bibfnamefont {L.~W.}\ \bibnamefont {Molenkamp}}, \bibinfo {author}
  {\bibfnamefont {T.}~\bibnamefont {Kiessling}}, \bibinfo {author}
  {\bibfnamefont {F.}~\bibnamefont {Schindler}}, \bibinfo {author}
  {\bibfnamefont {C.~H.}\ \bibnamefont {Lee}}, \bibinfo {author} {\bibfnamefont
  {M.}~\bibnamefont {Greiter}}, \bibinfo {author} {\bibfnamefont
  {T.}~\bibnamefont {Neupert}}, \ and\ \bibinfo {author} {\bibfnamefont
  {R.}~\bibnamefont {Thomale}},\ }\href {\doibase 10.1038/s41567-018-0246-1}
  {\bibfield  {journal} {\bibinfo  {journal} {Nature Physics}\ }\textbf
  {\bibinfo {volume} {14}},\ \bibinfo {pages} {925} (\bibinfo {year}
  {2018})}\BibitemShut {NoStop}%
\bibitem [{\citenamefont {Ozawa}\ \emph {et~al.}(2019)\citenamefont {Ozawa},
  \citenamefont {Price}, \citenamefont {Amo}, \citenamefont {Goldman},
  \citenamefont {Hafezi}, \citenamefont {Lu}, \citenamefont {Rechtsman},
  \citenamefont {Schuster}, \citenamefont {Simon}, \citenamefont {Zilberberg},\
  and\ \citenamefont {Carusotto}}]{Ozawa2019RMP}%
  \BibitemOpen
  \bibfield  {author} {\bibinfo {author} {\bibfnamefont {T.}~\bibnamefont
  {Ozawa}}, \bibinfo {author} {\bibfnamefont {H.~M.}\ \bibnamefont {Price}},
  \bibinfo {author} {\bibfnamefont {A.}~\bibnamefont {Amo}}, \bibinfo {author}
  {\bibfnamefont {N.}~\bibnamefont {Goldman}}, \bibinfo {author} {\bibfnamefont
  {M.}~\bibnamefont {Hafezi}}, \bibinfo {author} {\bibfnamefont
  {L.}~\bibnamefont {Lu}}, \bibinfo {author} {\bibfnamefont {M.~C.}\
  \bibnamefont {Rechtsman}}, \bibinfo {author} {\bibfnamefont {D.}~\bibnamefont
  {Schuster}}, \bibinfo {author} {\bibfnamefont {J.}~\bibnamefont {Simon}},
  \bibinfo {author} {\bibfnamefont {O.}~\bibnamefont {Zilberberg}}, \ and\
  \bibinfo {author} {\bibfnamefont {I.}~\bibnamefont {Carusotto}},\ }\href
  {\doibase 10.1103/RevModPhys.91.015006} {\bibfield  {journal} {\bibinfo
  {journal} {Rev. Mod. Phys.}\ }\textbf {\bibinfo {volume} {91}},\ \bibinfo
  {pages} {015006} (\bibinfo {year} {2019})}\BibitemShut {NoStop}%
\bibitem [{\citenamefont {Ma}\ \emph {et~al.}(2019)\citenamefont {Ma},
  \citenamefont {Xiao},\ and\ \citenamefont {Chan}}]{MaGC_2019nature}%
  \BibitemOpen
  \bibfield  {author} {\bibinfo {author} {\bibfnamefont {G.}~\bibnamefont
  {Ma}}, \bibinfo {author} {\bibfnamefont {M.}~\bibnamefont {Xiao}}, \ and\
  \bibinfo {author} {\bibfnamefont {C.~T.}\ \bibnamefont {Chan}},\ }\href
  {\doibase 10.1038/s42254-019-0030-x} {\bibfield  {journal} {\bibinfo
  {journal} {Nature Reviews Physics}\ }\textbf {\bibinfo {volume} {1}},\
  \bibinfo {pages} {281} (\bibinfo {year} {2019})}\BibitemShut {NoStop}%
\bibitem [{\citenamefont {Serra-Garcia}\ \emph {et~al.}(2018)\citenamefont
  {Serra-Garcia}, \citenamefont {Peri}, \citenamefont {Süsstrunk},
  \citenamefont {Bilal}, \citenamefont {Larsen}, \citenamefont {Villanueva},\
  and\ \citenamefont {Huber}}]{Serra_Garcia_2018nature}%
  \BibitemOpen
  \bibfield  {author} {\bibinfo {author} {\bibfnamefont {M.}~\bibnamefont
  {Serra-Garcia}}, \bibinfo {author} {\bibfnamefont {V.}~\bibnamefont {Peri}},
  \bibinfo {author} {\bibfnamefont {R.}~\bibnamefont {Süsstrunk}}, \bibinfo
  {author} {\bibfnamefont {O.~R.}\ \bibnamefont {Bilal}}, \bibinfo {author}
  {\bibfnamefont {T.}~\bibnamefont {Larsen}}, \bibinfo {author} {\bibfnamefont
  {L.~G.}\ \bibnamefont {Villanueva}}, \ and\ \bibinfo {author} {\bibfnamefont
  {S.~D.}\ \bibnamefont {Huber}},\ }\href {\doibase 10.1038/nature25156}
  {\bibfield  {journal} {\bibinfo  {journal} {Nature}\ }\textbf {\bibinfo
  {volume} {555}},\ \bibinfo {pages} {342} (\bibinfo {year}
  {2018})}\BibitemShut {NoStop}%
\bibitem [{\citenamefont {Yu}\ \emph {et~al.}(2020)\citenamefont {Yu},
  \citenamefont {Zhao},\ and\ \citenamefont {Schnyder}}]{Yu_Zhao_NSR}%
  \BibitemOpen
  \bibfield  {author} {\bibinfo {author} {\bibfnamefont {R.}~\bibnamefont
  {Yu}}, \bibinfo {author} {\bibfnamefont {Y.~X.}\ \bibnamefont {Zhao}}, \ and\
  \bibinfo {author} {\bibfnamefont {A.~P.}\ \bibnamefont {Schnyder}},\ }\href
  {\doibase 10.1093/nsr/nwaa065} {\bibfield  {journal} {\bibinfo  {journal}
  {National Science Review}\ } (\bibinfo {year} {2020}),\
  10.1093/nsr/nwaa065}\BibitemShut {NoStop}%
\bibitem [{\citenamefont {Peterson}\ \emph {et~al.}(2018)\citenamefont
  {Peterson}, \citenamefont {Benalcazar}, \citenamefont {Hughes},\ and\
  \citenamefont {Bahl}}]{Peterson_2018nature}%
  \BibitemOpen
  \bibfield  {author} {\bibinfo {author} {\bibfnamefont {C.~W.}\ \bibnamefont
  {Peterson}}, \bibinfo {author} {\bibfnamefont {W.~A.}\ \bibnamefont
  {Benalcazar}}, \bibinfo {author} {\bibfnamefont {T.~L.}\ \bibnamefont
  {Hughes}}, \ and\ \bibinfo {author} {\bibfnamefont {G.}~\bibnamefont
  {Bahl}},\ }\href {\doibase 10.1038/nature25777} {\bibfield  {journal}
  {\bibinfo  {journal} {Nature}\ }\textbf {\bibinfo {volume} {555}},\ \bibinfo
  {pages} {346} (\bibinfo {year} {2018})}\BibitemShut {NoStop}%
\bibitem [{\citenamefont {Yu}\ \emph {et~al.}(2015)\citenamefont {Yu},
  \citenamefont {Weng}, \citenamefont {Fang}, \citenamefont {Dai},\ and\
  \citenamefont {Hu}}]{YuRui2015prl}%
  \BibitemOpen
  \bibfield  {author} {\bibinfo {author} {\bibfnamefont {R.}~\bibnamefont
  {Yu}}, \bibinfo {author} {\bibfnamefont {H.}~\bibnamefont {Weng}}, \bibinfo
  {author} {\bibfnamefont {Z.}~\bibnamefont {Fang}}, \bibinfo {author}
  {\bibfnamefont {X.}~\bibnamefont {Dai}}, \ and\ \bibinfo {author}
  {\bibfnamefont {X.}~\bibnamefont {Hu}},\ }\href {\doibase
  10.1103/PhysRevLett.115.036807} {\bibfield  {journal} {\bibinfo  {journal}
  {Phys. Rev. Lett.}\ }\textbf {\bibinfo {volume} {115}},\ \bibinfo {pages}
  {036807} (\bibinfo {year} {2015})}\BibitemShut {NoStop}%
\bibitem [{\citenamefont {Sheng}\ \emph {et~al.}(2019)\citenamefont {Sheng},
  \citenamefont {Chen}, \citenamefont {Liu}, \citenamefont {Chen},
  \citenamefont {Yu}, \citenamefont {Zhao},\ and\ \citenamefont
  {Yang}}]{ZhaoYang19prl}%
  \BibitemOpen
  \bibfield  {author} {\bibinfo {author} {\bibfnamefont {X.-L.}\ \bibnamefont
  {Sheng}}, \bibinfo {author} {\bibfnamefont {C.}~\bibnamefont {Chen}},
  \bibinfo {author} {\bibfnamefont {H.}~\bibnamefont {Liu}}, \bibinfo {author}
  {\bibfnamefont {Z.}~\bibnamefont {Chen}}, \bibinfo {author} {\bibfnamefont
  {Z.-M.}\ \bibnamefont {Yu}}, \bibinfo {author} {\bibfnamefont {Y.~X.}\
  \bibnamefont {Zhao}}, \ and\ \bibinfo {author} {\bibfnamefont {S.~A.}\
  \bibnamefont {Yang}},\ }\href {\doibase 10.1103/PhysRevLett.123.256402}
  {\bibfield  {journal} {\bibinfo  {journal} {Phys. Rev. Lett.}\ }\textbf
  {\bibinfo {volume} {123}},\ \bibinfo {pages} {256402} (\bibinfo {year}
  {2019})}\BibitemShut {NoStop}%
\bibitem [{\citenamefont {Wu}\ \emph {et~al.}(2019)\citenamefont {Wu},
  \citenamefont {Soluyanov},\ and\ \citenamefont {Bzdu{\v s}ek}}]{Wu1273}%
  \BibitemOpen
  \bibfield  {author} {\bibinfo {author} {\bibfnamefont {Q.}~\bibnamefont
  {Wu}}, \bibinfo {author} {\bibfnamefont {A.~A.}\ \bibnamefont {Soluyanov}}, \
  and\ \bibinfo {author} {\bibfnamefont {T.}~\bibnamefont {Bzdu{\v s}ek}},\
  }\href {\doibase 10.1126/science.aau8740} {\bibfield  {journal} {\bibinfo
  {journal} {Science}\ }\textbf {\bibinfo {volume} {365}},\ \bibinfo {pages}
  {1273} (\bibinfo {year} {2019})}\BibitemShut {NoStop}%
\bibitem [{\citenamefont {Wang}\ \emph {et~al.}(2019)\citenamefont {Wang},
  \citenamefont {Wieder}, \citenamefont {Li}, \citenamefont {Yun},\ and\
  \citenamefont {Bernevig}}]{Wangzhijun2019prl}%
  \BibitemOpen
  \bibfield  {author} {\bibinfo {author} {\bibfnamefont {Z.}~\bibnamefont
  {Wang}}, \bibinfo {author} {\bibfnamefont {B.~J.}\ \bibnamefont {Wieder}},
  \bibinfo {author} {\bibfnamefont {J.}~\bibnamefont {Li}}, \bibinfo {author}
  {\bibfnamefont {B.}~\bibnamefont {Yun}}, \ and\ \bibinfo {author}
  {\bibfnamefont {B.~A.}\ \bibnamefont {Bernevig}},\ }\href {\doibase
  10.1103/PhysRevLett.123.186401} {\bibfield  {journal} {\bibinfo  {journal}
  {Phys. Rev. Lett.}\ }\textbf {\bibinfo {volume} {123}},\ \bibinfo {pages}
  {186401} (\bibinfo {year} {2019})}\BibitemShut {NoStop}%
\bibitem [{\citenamefont {Li}\ \emph {et~al.}(2020)\citenamefont {Li},
  \citenamefont {Mekawy}, \citenamefont {Krasnok},\ and\ \citenamefont
  {Al\`u}}]{LiHN2020prl}%
  \BibitemOpen
  \bibfield  {author} {\bibinfo {author} {\bibfnamefont {H.}~\bibnamefont
  {Li}}, \bibinfo {author} {\bibfnamefont {A.}~\bibnamefont {Mekawy}}, \bibinfo
  {author} {\bibfnamefont {A.}~\bibnamefont {Krasnok}}, \ and\ \bibinfo
  {author} {\bibfnamefont {A.}~\bibnamefont {Al\`u}},\ }\href {\doibase
  10.1103/PhysRevLett.124.193901} {\bibfield  {journal} {\bibinfo  {journal}
  {Phys. Rev. Lett.}\ }\textbf {\bibinfo {volume} {124}},\ \bibinfo {pages}
  {193901} (\bibinfo {year} {2020})}\BibitemShut {NoStop}%
\bibitem [{\citenamefont {Wang}\ \emph {et~al.}(2020)\citenamefont {Wang},
  \citenamefont {Dai}, \citenamefont {Shao}, \citenamefont {Yang},\ and\
  \citenamefont {Zhao}}]{Wang2020}%
  \BibitemOpen
  \bibfield  {author} {\bibinfo {author} {\bibfnamefont {K.}~\bibnamefont
  {Wang}}, \bibinfo {author} {\bibfnamefont {J.-X.}\ \bibnamefont {Dai}},
  \bibinfo {author} {\bibfnamefont {L.~B.}\ \bibnamefont {Shao}}, \bibinfo
  {author} {\bibfnamefont {S.~A.}\ \bibnamefont {Yang}}, \ and\ \bibinfo
  {author} {\bibfnamefont {Y.~X.}\ \bibnamefont {Zhao}},\ }\href {\doibase
  10.1103/PhysRevLett.125.126403} {\bibfield  {journal} {\bibinfo  {journal}
  {Phys. Rev. Lett.}\ }\textbf {\bibinfo {volume} {125}},\ \bibinfo {pages}
  {126403} (\bibinfo {year} {2020})}\BibitemShut {NoStop}%
\bibitem [{\citenamefont {Chen}\ \emph {et~al.}(2021)\citenamefont {Chen},
  \citenamefont {Wu}, \citenamefont {Yu}, \citenamefont {Chen}, \citenamefont
  {Zhao}, \citenamefont {Sheng},\ and\ \citenamefont
  {Yang}}]{chen2021graphyne}%
  \BibitemOpen
  \bibfield  {author} {\bibinfo {author} {\bibfnamefont {C.}~\bibnamefont
  {Chen}}, \bibinfo {author} {\bibfnamefont {W.}~\bibnamefont {Wu}}, \bibinfo
  {author} {\bibfnamefont {Z.-M.}\ \bibnamefont {Yu}}, \bibinfo {author}
  {\bibfnamefont {Z.}~\bibnamefont {Chen}}, \bibinfo {author} {\bibfnamefont
  {Y.~X.}\ \bibnamefont {Zhao}}, \bibinfo {author} {\bibfnamefont {X.-L.}\
  \bibnamefont {Sheng}}, \ and\ \bibinfo {author} {\bibfnamefont {S.~A.}\
  \bibnamefont {Yang}},\ }\href {\doibase 10.1103/PhysRevB.104.085205}
  {\bibfield  {journal} {\bibinfo  {journal} {Phys. Rev. B}\ }\textbf {\bibinfo
  {volume} {104}},\ \bibinfo {pages} {085205} (\bibinfo {year}
  {2021})}\BibitemShut {NoStop}%
\bibitem [{\citenamefont {Chen}\ \emph {et~al.}(2022)\citenamefont {Chen},
  \citenamefont {Zeng}, \citenamefont {Chen}, \citenamefont {Zhao},
  \citenamefont {Sheng},\ and\ \citenamefont {Yang}}]{chen2022second}%
  \BibitemOpen
  \bibfield  {author} {\bibinfo {author} {\bibfnamefont {C.}~\bibnamefont
  {Chen}}, \bibinfo {author} {\bibfnamefont {X.-T.}\ \bibnamefont {Zeng}},
  \bibinfo {author} {\bibfnamefont {Z.}~\bibnamefont {Chen}}, \bibinfo {author}
  {\bibfnamefont {Y.~X.}\ \bibnamefont {Zhao}}, \bibinfo {author}
  {\bibfnamefont {X.-L.}\ \bibnamefont {Sheng}}, \ and\ \bibinfo {author}
  {\bibfnamefont {S.~A.}\ \bibnamefont {Yang}},\ }\href {\doibase
  10.1103/PhysRevLett.128.026405} {\bibfield  {journal} {\bibinfo  {journal}
  {Phys. Rev. Lett.}\ }\textbf {\bibinfo {volume} {128}},\ \bibinfo {pages}
  {026405} (\bibinfo {year} {2022})}\BibitemShut {NoStop}%
\bibitem [{\citenamefont {Dai}\ \emph {et~al.}(2021)\citenamefont {Dai},
  \citenamefont {Wang}, \citenamefont {Yang},\ and\ \citenamefont
  {Zhao}}]{Dai2020}%
  \BibitemOpen
  \bibfield  {author} {\bibinfo {author} {\bibfnamefont {J.-X.}\ \bibnamefont
  {Dai}}, \bibinfo {author} {\bibfnamefont {K.}~\bibnamefont {Wang}}, \bibinfo
  {author} {\bibfnamefont {S.~A.}\ \bibnamefont {Yang}}, \ and\ \bibinfo
  {author} {\bibfnamefont {Y.~X.}\ \bibnamefont {Zhao}},\ }\href {\doibase
  10.1103/PhysRevB.104.165142} {\bibfield  {journal} {\bibinfo  {journal}
  {Phys. Rev. B}\ }\textbf {\bibinfo {volume} {104}},\ \bibinfo {pages}
  {165142} (\bibinfo {year} {2021})}\BibitemShut {NoStop}%
\bibitem [{Gam()}]{Gamma-Point}%
  \BibitemOpen
  \href@noop {} {}\bibinfo {note} {By the numerical calculation of Wilson loop,
  one can easily check that only the algebra of parameters obtained at $\Gamma$
  point is the real bulk criticality, while the others are not.}\BibitemShut
  {Stop}%
\bibitem [{\citenamefont {Haldane}(1988)}]{Haldane1988prl}%
  \BibitemOpen
  \bibfield  {author} {\bibinfo {author} {\bibfnamefont {F.~D.~M.}\
  \bibnamefont {Haldane}},\ }\href {\doibase 10.1103/PhysRevLett.61.2015}
  {\bibfield  {journal} {\bibinfo  {journal} {Phys. Rev. Lett.}\ }\textbf
  {\bibinfo {volume} {61}},\ \bibinfo {pages} {2015} (\bibinfo {year}
  {1988})}\BibitemShut {NoStop}%
\bibitem [{Edg()}]{Edge-Mass}%
  \BibitemOpen
  \href@noop {} {}\bibinfo {note} {Note that a pair of opposite edges parallel
  to $\mathbf{a}_i$ have opposite mass terms since $\P\T$ symmetry inverses the
  effective mass.}\BibitemShut {Stop}%
\bibitem [{Fin()}]{Finite-size}%
  \BibitemOpen
  \href@noop {} {}\bibinfo {note} {Finite-size effects can split the degenerate
  zero modes and deviate them from zero to form a ingap corner modes, but the
  deviation is exponentially suppressed with the sample size.}\BibitemShut
  {Stop}%
\bibitem [{\citenamefont {Jackiw}\ and\ \citenamefont
  {Rebbi}(1976)}]{Jackiw1976prd}%
  \BibitemOpen
  \bibfield  {author} {\bibinfo {author} {\bibfnamefont {R.}~\bibnamefont
  {Jackiw}}\ and\ \bibinfo {author} {\bibfnamefont {C.}~\bibnamefont {Rebbi}},\
  }\href {\doibase 10.1103/PhysRevD.13.3398} {\bibfield  {journal} {\bibinfo
  {journal} {Phys. Rev. D}\ }\textbf {\bibinfo {volume} {13}},\ \bibinfo
  {pages} {3398} (\bibinfo {year} {1976})}\BibitemShut {NoStop}%
\bibitem [{\citenamefont {Yang}\ \emph {et~al.}(2020)\citenamefont {Yang},
  \citenamefont {Li}, \citenamefont {Peng}, \citenamefont {Zou},\ and\
  \citenamefont {Cheng}}]{PhysRevLett.125.255502}%
  \BibitemOpen
  \bibfield  {author} {\bibinfo {author} {\bibfnamefont {Z.-Z.}\ \bibnamefont
  {Yang}}, \bibinfo {author} {\bibfnamefont {X.}~\bibnamefont {Li}}, \bibinfo
  {author} {\bibfnamefont {Y.-Y.}\ \bibnamefont {Peng}}, \bibinfo {author}
  {\bibfnamefont {X.-Y.}\ \bibnamefont {Zou}}, \ and\ \bibinfo {author}
  {\bibfnamefont {J.-C.}\ \bibnamefont {Cheng}},\ }\href {\doibase
  10.1103/PhysRevLett.125.255502} {\bibfield  {journal} {\bibinfo  {journal}
  {Phys. Rev. Lett.}\ }\textbf {\bibinfo {volume} {125}},\ \bibinfo {pages}
  {255502} (\bibinfo {year} {2020})}\BibitemShut {NoStop}%
\bibitem [{\citenamefont {Noh}\ \emph {et~al.}(2018)\citenamefont {Noh},
  \citenamefont {Benalcazar}, \citenamefont {Huang}, \citenamefont {Collins},
  \citenamefont {Chen}, \citenamefont {Hughes},\ and\ \citenamefont
  {Rechtsman}}]{photonic}%
  \BibitemOpen
  \bibfield  {author} {\bibinfo {author} {\bibfnamefont {J.}~\bibnamefont
  {Noh}}, \bibinfo {author} {\bibfnamefont {W.~A.}\ \bibnamefont {Benalcazar}},
  \bibinfo {author} {\bibfnamefont {S.}~\bibnamefont {Huang}}, \bibinfo
  {author} {\bibfnamefont {M.~J.}\ \bibnamefont {Collins}}, \bibinfo {author}
  {\bibfnamefont {K.~P.}\ \bibnamefont {Chen}}, \bibinfo {author}
  {\bibfnamefont {T.~L.}\ \bibnamefont {Hughes}}, \ and\ \bibinfo {author}
  {\bibfnamefont {M.~C.}\ \bibnamefont {Rechtsman}},\ }\href {\doibase
  10.1038/s41566-018-0179-3} {\bibfield  {journal} {\bibinfo  {journal} {Nature
  Photon}\ }\textbf {\bibinfo {volume} {12}},\ \bibinfo {pages} {408–415}
  (\bibinfo {year} {2018})}\BibitemShut {NoStop}%
\end{thebibliography}%

\end{document}